\documentclass[a4paper,12pt]{article}

\usepackage{amsmath,amssymb}
\usepackage{subfigure}
\usepackage{graphicx}
\usepackage[numbers,sort&compress]{natbib}
\usepackage{xcolor}
\usepackage[colorlinks,
            linkcolor=black,
            filecolor=black,
            anchorcolor=black,
            urlcolor=black,
            citecolor=black
            ]{hyperref}
\usepackage{ulem}
\usepackage{setspace}
\usepackage{array}
\usepackage{fancyhdr}
\usepackage{bm}

\newlength{\dinwidth}
\newlength{\dinmargin}
\fancypagestyle{firstpage}{
  \rhead{}
}

\newcommand{\SM}{{\rm SM}}
\newcommand{\NP}{{\rm NP}}

\newcommand{\GeV}{{\,\rm GeV}}
\newcommand{\TeV}{{\,\rm TeV}}

\newcommand{\fb}{{\, \rm fb}}

\newcommand{\mB}{{\mathcal B}}
\newcommand{\mO}{{\mathcal O}}
\newcommand{\mC}{{\mathcal C}}

\newcommand{\Zp}{{Z^\prime}}
\newcommand{\XctL}{{X_{ct}^L}}
\newcommand{\XctLII}{{X_{23}^L}}
\newcommand{\lmmL}{{\lambda_{\mu\mu}^L}}
\newcommand{\lmmLII}{{\lambda_{22}^L}}

\newcommand{\flavio}{{\texttt{flavio}}}

\newcommand{\re}{{\rm Re}}
\newcommand{\im}{{\rm Im}}

\allowdisplaybreaks

\begin{document}

\setlength{\abovedisplayskip}{6pt}
\setlength{\belowdisplayskip}{6pt}

\title{\bf \boldmath Explaining the $b \to s \ell^+ \ell^-$ anomalies \\ in $\Zp$ scenarios with top-FCNC couplings}

\author{
  Xin-Qiang Li${}^{1,}$\footnote{xqli@ccnu.edu.cn},\quad
  Meng Shen${}^1$\footnote{shenmeng@mails.ccnu.edu.cn},\quad
  Dong-Yang Wang${}^{2,}$\footnote{wangdongyang@mails.ccnu.edu.cn},\quad
  \\
  Ya-Dong Yang${}^{1,3,}$\footnote{yangyd@ccnu.edu.cn},\quad
  and\,
  Xing-Bo Yuan${}^{1,}$\footnote{y@ccnu.edu.cn}\\[15pt]
\small ${}^1$ Institute of Particle Physics and Key Laboratory of Quark and Lepton Physics~(MOE), \\[-0.2cm]
\small Central China Normal University, Wuhan, Hubei 430079, China\\[-0.2cm]
\small ${}^2$ Department of Physics, Jiujiang University, Jiujiang 332005, China\\[-0.2cm]
\small ${}^3$ School of Physics and Microelectronics, Zhengzhou University, Zhengzhou, Henan 450001, China
}

\date{}
\maketitle

\thispagestyle{firstpage}

\vspace{1em}

\begin{abstract}
Motivated by the recent anomalies in $b \to s \ell^+ \ell^-$ transitions, we explore a minimal $\Zp$ scenario, in which the $\Zp$ boson has a flavour-changing coupling to charm and top quarks and a flavour-conserving coupling to muons. It is found that such a $\Zp$ boson can explain the current $b \to s \ell^+ \ell^-$ anomalies, while satisfying other flavour and collider constraints simultaneously. The $\Zp$ boson can be as light as few hundreds GeV. In this case, the $t \to c \mu^+ \mu^-$ decay and the $t\Zp$ associated production at the LHC could provide sensitive probes of such a $\Zp$ boson. As a special feature, the $\Zp$ contributions to all rare $B$- and $K$-meson processes are controlled by one parameter. This results in interesting correlations among these processes, which could provide further insights into this scenario. In addition, an extended scenario, in which the $\Zp$ boson interacts with the $SU(2)_L$ fermion doublets with analogous couplings as in the minimal scenario, is also investigated. 
\end{abstract}

\newpage

\section{Introduction}
\label{sec:introduction}

The flavour-changing neutral current (FCNC) processes are sensitive to possible contributions from heavy mediators and provide a complementary way to direct searches for new physics (NP) at high-energy frontiers. While there is so far no direct evidence for NP at the LHC, recent measurements of the rare $b \to s \ell^+ \ell^-$ decays exhibit several interesting discrepancies from the Standard Model (SM) predictions of the branching ratios, the angular distributions, and the lepton flavour universality (LFU) ratios~\cite{Li:2018lxi,Bifani:2018zmi,Albrecht:2021tul,London:2021lfn}.

In this respect, the ratios defined as $R_{K^{(*)}} \equiv \mB(B \to K^{(*)} \mu^+ \mu^-)/\mB(B\to K^{(*)}e^+ e^-)$ are of particular interest, because the hadronic uncertainties largely cancel out here~\cite{Hiller:2003js}. Therefore, they provide a sensitive test of the LFU. In the SM, these ratios are predicted to be close to unity up to tiny electromagnetic corrections~\cite{Hiller:2003js,Bouchard:2013mia,Bordone:2016gaq,Isidori:2020acz}. Recently, the LHCb collaboration presented an updated measurement of $R_K$ using the full data set of $9\fb^{-1}$~\cite{Aaij:2021vac}: 
\begin{align}
  R_K = 0.846_{-0.039}^{+0.042}(\text{stat})_{-0.012}^{+0.013}(\text{syst}), \qquad \text{for }1.1<q^2<6.0 \GeV^2,
\end{align}
where $q^2$ denotes the dilepton invariant mass squared. This new result confirms the previous LHCb measurement using a data set of $5\fb^{-1}$~\cite{Aaij:2019wad}. However, the tension with the SM prediction has increased from previously $2.5\sigma$ to now $3.1\sigma$, due to reduced experimental uncertainties. LHCb has also reported a measurement of $R_{K^*}$ using the full Run-I data set of $3\fb^{-1}$~\cite{LHCb:2017avl}:
\begin{align}
  R_{K^*}=
  \begin{cases}
    0.66_{-0.07}^{+0.11}(\text{stat})\pm 0.03(\text{syst}), & 0.045 < q^2 < 1.1 \GeV^2,\\
    0.69_{-0.07}^{+0.11}(\text{stat})\pm 0.05(\text{syst}), & 1.1\hphantom{00} < q^2 < 6.0 \GeV^2,
  \end{cases}
\end{align}
which are found to be about $2.1\sigma$ and $2.4\sigma$ lower than the SM predictions, respectively. Very recently, the LHCb measurements of $R_{K_S}$ and $R_{K^{*+}}$ have also been reported~\cite{LHCb:2021lvy}. Previous measurements from the BaBar~\cite{BaBar:2012mrf} and Belle~\cite{Abdesselam:2019wac,BELLE:2019xld} experiments are also consistent with the LHCb results, although with relatively large uncertainties. All these measurements of $R_K$ and $R_{K^*}$ together provide intriguing hints of the LFU violation.

Recently, the LHCb collaboration reported the most precise measurement of the branching ratio of the $B_s \to \phi \mu^+ \mu^-$ decay using the full Run-1 and Run-2 data sets~\cite{LHCb:2021zwz},
\begin{align}
 \mB(B_s \to \phi \mu^+ \mu^-) = (2.88 \pm 0.15 \pm 0.05 \pm 0.14)\times 10^{-8},
  \quad \text{for }1.1<q^2<6.0 \GeV^2,
\end{align}
where the uncertainties are, in order, statistical, systematic, and from the branching fraction of the normalization mode. This measurement is found to lie $3.6\sigma$ below the SM prediction~\cite{Horgan:2013pva,Horgan:2015vla,Bharucha:2015bzk,Rajeev:2020aut}. Using the same full data set, LHCb later presented an improved measurement of the branching ratio of the rare $B_s \to \mu^+ \mu^-$ decay~\cite{LHCb:2021vsc},
\begin{align}
  \mB(B_s \to \mu^+ \mu^-)=  \big[3.09_{-0.43}^{+0.46}(\text{stat})_{-0.11}^{+0.15}(\text{syst}) \big]\times 10^{-9},
\end{align}
which is statistically consistent with the previous world average~\cite{LHCb:2020zud}. This new result is compatible with the SM expectation~\cite{Bobeth:2013uxa,Beneke:2019slt} within $1\sigma$. However, after combining all the measurements of the $B_{s,d} \to \mu^+ \mu^-$ decays from the ATLAS~\cite{ATLAS:2018cur}, CMS~\cite{CMS:2019bbr}, and LHCb~\cite{LHCb:2021vsc} experiments, the total discrepancy with the SM is found to be at the level of $2\sigma$~\cite{Altmannshofer:2021qrr,Geng:2021nhg}. In addition, the latest measurements of the $B^0 \to K^{*0} \mu^+ \mu^-$~\cite{LHCb:2020lmf} and $B^+ \to K^{*+} \mu^+ \mu^-$~\cite{LHCb:2020gog} decays show tensions in the angular distributions with respect to the SM predictions. The local discrepancies in the angular observables $P_2$ and $P_5^\prime$ are observed to be at $2.5$-$3.0\sigma$ in two $q^2$ bins~\cite{LHCb:2020lmf,LHCb:2020gog}.

Although none of the individual deviations is statistically significant, and further refinement of the hadronic uncertainties in some observables is still an ongoing theoretical issue~\cite{Ciuchini:2020gvn,Hurth:2020rzx,Gubernari:2020eft}, the global tension in the $b \to s \ell^+ \ell^-$ decays has motivated a lot of NP interpretations~\cite{Li:2018lxi,Bifani:2018zmi,Albrecht:2021tul,London:2021lfn}. In this respect, one of the most popular NP explanations are models with an extra heavy vector $\Zp$ boson. In these models, the $\Zp$ boson has couplings to quarks, as well as to either electrons or muons. Depending on the quark couplings involved, these $\Zp$ models can be classified into two categories: (i) The $\Zp$ boson has flavour-violating couplings to $b$ and $s$ quarks and the $b \to s \ell^+ \ell^-$ transitions receive contributions from tree-level $\Zp$ exchange~\cite{Gauld:2013qba,Buras:2013qja,Altmannshofer:2014cfa,Crivellin:2015mga,Crivellin:2015lwa,Crivellin:2015era,Celis:2015ara,Belanger:2015nma,Falkowski:2015zwa,Allanach:2015gkd,Ko:2017lzd,DiChiara:2017cjq,Bonilla:2017lsq,Ellis:2017nrp,Tang:2017gkz,King:2017anf,Baek:2017sew,King:2018fcg,Allanach:2018odd,Allanach:2019iiy,Altmannshofer:2019xda,Calibbi:2019lvs,Mohapatra:2021ynn,Lozano:2021zbu,Rajeev:2021ntt}. (ii) The $\Zp$ boson has flavour-conserving couplings to top quark and affects the $b \to s \ell^+ \ell^-$ transitions via one-loop penguin diagrams~\cite{Kamenik:2017tnu,Fox:2018ldq}. In this paper, based on our previous works~\cite{Li:2011af,Gong:2013sh}, we will consider another possibility, where the $\Zp$ boson has flavour-violating couplings to top and charm quarks. This scenario does not suffer from the constraints from $B_s - \bar B_s$ mixing, and the $\Zp$ boson contributes to the $b \to s \ell^+ \ell^-$ decays at the one-loop level. We will derive constraints on the $\Zp$ mass and couplings from various flavour and collider processes, and study the possibility of explaining the $b \to s \ell^+ \ell^-$ anomalies in such a scenario. Future prospects of searching for such a $\Zp$ boson at the LHC will also be discussed.

This paper is organized as follows. In section~\ref{sec:Zp}, we introduce the phenomenological $\Zp$ scenarios, which have the desired flavour-changing couplings to explain the current $b \to s \ell^+ \ell^-$ anomalies. In section~\ref{sec:framework}, we recapitulate the theoretical frameworks for various flavour processes and discuss the $\Zp$ effects. In section~\ref{sec:num}, we give our detailed numerical results and discussions. Our conclusions are finally made in section~\ref{sec:conclusion}.

\section{$\Zp$ scenarios with top-quark FCNC couplings}
\label{sec:Zp}

We consider two phenomenological scenarios involving a $\Zp$ boson. In the first scenario (denoted as scenario I), the $\Zp$ boson has flavour-changing couplings to $c$ and $t$ quarks and a flavour-conserving coupling to the $\mu$ lepton. Their interactions are described by the effective Lagrangian
\begin{align}\label{eq:LZp:I}
\mathcal L_\Zp^{\rm I}= \big(\XctL\, \bar c \gamma^\mu P_L t\, Z^\prime_\mu +{\rm h.c.} \big) +\lmmL\, \bar{\mu}\gamma^\mu P_L\mu\, Z^\prime_\mu,
\end{align}
where $P_L=(1-\gamma_5)/2$, and the fermion fields $c$, $t$ and $\mu$ refer to the mass eigenstates. Generally, the couplings $\XctL$ and $\lmmL$ are complex and real, respectively. In this scenario, the $\Zp$ boson affects the $b \to s \mu^+ \mu^-$ transitions via one-loop penguin diagram and could explain the anomalies observed in $B \to K^{(*)} \ell^+ \ell^-$ decays. Furthermore, sizable contributions to the $s \to d \mu^+ \mu^-$ transitions could arise from the $\Zp$ penguin diagram.

In the second scenario (denoted as scenario II), the $\Zp$ boson is assumed to interact with the $SU(2)_L$ fermion doublets with similar couplings as in scenario I. In the mass eigenstate basis, the effective Lagrangian takes the form\footnote{In ref.~\cite{Coy:2019rfr}, a similar $\Zp$ scenario has been discussed, where the $\Zp$ boson has left-handed couplings to the lepton doublets but only right-handed couplings to the up-type quarks.}
\begin{align}\label{eq:LZp:II}
  \mathcal L_\Zp^{\rm II}=& \big(X_{23}^L\,\bar Q_{L,2}\gamma^\mu Q_{L,3}\, Z_\mu^\prime + \text{h.c.} \big) +\lambda^L_{22}\,\bar L_{L,2} \gamma^\mu L_{L,2}\, Z_\mu^\prime
  \\
  =& \big [X_{23}^L \big (\bar c\gamma^\mu P_L t+\bar s \gamma^\mu P_L b \big ) Z_\mu^\prime + \text{h.c.} \big] + \lambda^L_{22} \big (\bar\mu \gamma^\mu P_L\mu+\bar\nu_\mu \gamma^\mu P_L\nu_\mu \big ) Z_\mu^\prime,\nonumber
\end{align}
where $Q_{L,i}$ and $L_{L,i}$ denote the left-handed $SU(2)_L$ quark and lepton doublets of the $i$-th generation, respectively. As in scenario I, the couplings $X_{23}^L$ and $\lambda_{22}^L$ are generally complex and real, respectively. In this scenario, the $b \bar s \Zp$ and the $t \bar c \Zp$ interaction have the same coupling strength due to $SU(2)_L$ invariance. As a consequence, the $\Zp$ contribution to $b \to s$ processes is mainly induced by the $b \bar s\Zp$ interaction at the tree level, and the phenomenology of these processes is similar to that in the so-called minimal $\Zp$ scenario discussed in refs.~\cite{Gauld:2013qba,DiChiara:2017cjq}. As in scenario I but contrary to the minimal $\Zp$ scenario, the $s \to d$ transitions can receive the $\Zp$ contribution at the one-loop level. Furthermore, the $\Zp$ boson also couples to the muon neutrino due to $SU(2)_L$ invariance, which may affect the $b \to s \nu \bar\nu$ and $s \to d \nu \bar\nu$ processes. The $\Zp$ effects in various flavour processes will be discussed in detail in the next section.

Alternatively, one can consider a $\Zp$ boson with flavour-changing couplings to $u$ and $t$ instead of $c$ and $t$ quarks in the above scenarios. Then, the $\Zp$ contributions to the top-quark production and FCNC decays can be very different from that in scenarios I and II. For the $\Zp$ couplings to leptons, a right-handed $\mu^+\mu^-\Zp$ interaction can also be added to simultaneously accommodate the $b \to s \mu^+ \mu^-$ and $(g-2)_\mu$ anomalies~\cite{Muong-2:2021ojo,Aoyama:2020ynm}. Similarly, scenarios with an $e^+e^-\Zp$ instead of a $\mu^+\mu^- \Zp$ coupling in the above scenarios can also be considered. Such a $\Zp$ boson could be directly produced in $e^+ e^-$ colliders, but loses the possibility of explaining the $(g-2)_\mu$ anomaly. We leave all these possibilities for future studies, and consider in this paper only the above two scenarios that have the couplings required to explain the $b\to s\ell^+\ell^-$ anomalies.

\section{$\Zp$ effects in various flavour processes}
\label{sec:framework}

In this section, we recapitulate the theoretical frameworks for various low-energy flavour processes and discuss the $\Zp$ contributions to them as well as to the top-quark physics.

\subsection{$b \to s \mu^+ \mu^-$ transitions}

\begin{figure}[t]
	\centering
	\subfigure[]{\hspace{-0.8em}\includegraphics[width=0.24\textwidth]{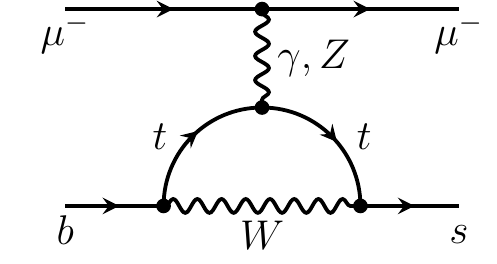}}
	\qquad
	\subfigure[]{\hspace{-0.8em}\includegraphics[width=0.24\textwidth]{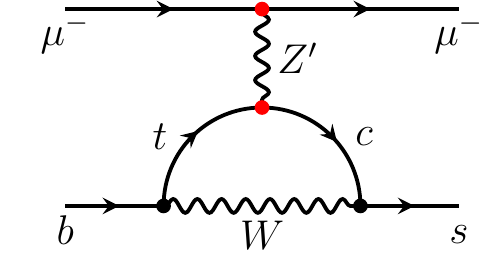}}
	\qquad
	\subfigure[]{\hspace{-0.8em}\includegraphics[width=0.24\textwidth]{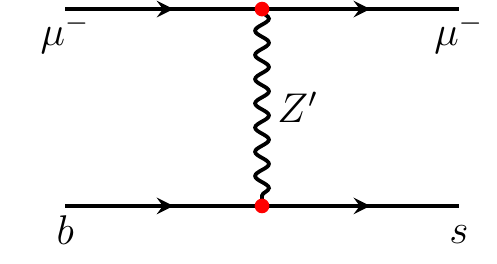}}
	\caption{Feynman diagrams for the $b \to s \mu^+ \mu^-$ transitions, including the selected SM penguin contributions (a), as well as the $\Zp$ effects from the scenarios I (b) and II (b, c).}
	\label{fig:b2smumu}
\end{figure}

The rare decays $ B_{s} \to \mu^+ \mu^- $, $B \to K^{(*)}\mu^+ \mu^-$, and $B_s \to \phi \mu^+\mu^-$ are induced by the $b \to s \mu^+ \mu^-$ transitions, and provide promising probes of NP effects. With the $\Zp$ contributions taken into account,  the effective Hamiltonian for the $b\to s\mu^+\mu^-$ transitions can be written as~\cite{Buchalla:1995vs}
\begin{align}\label{eq:Heff}
\mathcal H_{\rm eff}^{b \to s \mu^+ \mu^-}=-\frac{4G_F}{\sqrt{2}} V_{tb}V_{ts}^*\sum_{i=1}^{10} \mC_i  \mO_i,
\end{align}
where explicit definitions of the effective operators $\mO_{1-8}$ can be found in ref.~\cite{Buchalla:1995vs}. The operators most relevant to our study are, however, the two semi-leptonic operators $\mO_{9\ell}$ and $\mO_{10\ell}$ defined by
\begin{align}
  \mO_{9\ell} = \frac{e^2}{16 \pi^2}(\bar{s} \gamma_{\mu} P_L b)(\bar{\ell} \gamma^\mu \ell),
  \qquad \text{and} \qquad
\mO_{10\ell} =\frac{e^2}{16 \pi^2}(\bar{s}  \gamma_{\mu} P_L b)(  \bar{\ell} \gamma^\mu \gamma_5 \ell),
\end{align}
respectively. In the SM, their Wilson coefficients have been calculated including next-to-next-to-leading-order (NNLO) QCD~\cite{Bobeth:2003at,Huber:2005ig,Hermann:2013kca} and next-to-leading-order (NLO) electroweak corrections~\cite{Bobeth:2013tba}. In the $\Zp$ scenario I, the $t\bar{c}\Zp$ vertex can affect the $b\to s\mu^+\mu^-$ transitions through the $\Zp$-penguin diagram shown in figure~\ref{fig:b2smumu}\,(b), with the resulting NP Wilson coefficients given by~\cite{Li:2011af,Gong:2013sh}
\begin{align}
  \mC_{9\mu}^{\rm NP,\,I} = - \mC_{10\mu}^{\rm NP,\,I} = \frac{1}{8\sqrt{2}G_F s_W^2}\frac{V_{cs}^*}{V_{ts}^*}\frac{\XctL \lmmL }{m_\Zp^2}f(x_t),
\end{align}
with $s_W=\sin\theta_W$, $x_t = \bar{m}_t(\bar{m}_t)^2 / m_W^2$, and $V_{ij}$ the Cabibbo-Kobayashi-Maskawa (CKM) matrix elements~\cite{Cabibbo:1963yz,Kobayashi:1973fv}. The loop function $f(x)$ is defined as
\begin{align}\label{eq:loop}
  f(x) = \frac{3}{2}+x \log\frac{\mu^2}{m_W^2}+x-x \log x,
\end{align}
which is obtained from the calculation of similar diagrams with anomalous $t\bar{c}Z$ vertex in refs.~\cite{Li:2011af,Gong:2013sh}. It is noted that the $\Zp$ contributions are enhanced by the CKM factor $V_{cs}/V_{ts}$. In the $\Zp$ scenario II, the $\Zp$ boson can also affect the $b \to s \mu^+ \mu^-$ transitions at tree level through the diagram in figure~\ref{fig:b2smumu}\,(c). The total contributions to the NP Wilson coefficients are given by
\begin{align}
  \mC_{9\mu}^{\rm NP,\, II} = - \mC_{10\mu}^{\rm NP,\, II} = \frac{1}{\sqrt{2} G_F} \bigg(\frac{1}{8s_W^2}\frac{V_{cs}^*}{V_{ts}^*}f(x_t) - \frac{\pi}{\alpha_{\rm e}V_{tb}V_{ts}^*}\bigg)\frac{X_{23}^L\lambda_{22}^L}{m_{\Zp}^2}.
\end{align}
Numerically, it is found that the tree-level contribution is dominant and the loop-level contribution can be safely neglected. For $m_\Zp=1\,\TeV$, $\mC_{9\mu}^{\rm NP,\,I}=(-4.4-0.1 i) \XctL \lmmL$ in scenario I, while $\mC_{9\mu}^{\rm NP,\, II} = (585+11i) X_{23}^L\lambda_{22}^L$ in scenario II.

A lot of effort has been put into the theoretical treatment of $B \to K \mu^+ \mu^-$, $B\to K^{*}\mu^+\mu^-$, and $B_s \to \phi \mu^+\mu^-$  decays~\cite{Beneke:2004dp,Khodjamirian:2010vf,Beylich:2011aq,Bobeth:2012vn,Jager:2012uw,Bharucha:2015bzk,Gubernari:2020eft}. Besides the LFU ratios $R_{K^{(*)}}$ introduced in section~\ref{sec:introduction}, the angular observables in these decays are also known to provide valuable information of potential NP contributions, and hence have been analyzed in detail in refs.~\cite{Altmannshofer:2008dz,Descotes-Genon:2013vna}. For these observables, the main theoretical uncertainties come from the heavy-to-light transition form factors. During recent years, significant progress has been made in calculating these form factors from lattice QCD~\cite{Horgan:2015vla,Aoki:2021kgd} and light-cone sum rules (LCSR)~\cite{Ball:2004rg,Bharucha:2015bzk,Gubernari:2018wyi,Lu:2018cfc,Gao:2019lta,Shen:2021yhe}. For a recent review of the $b \to s \mu^+ \mu^-$ decays, we refer to refs.~\cite{Altmannshofer:2014rta,Descotes-Genon:2015uva,Blake:2016olu}, where the SM calculations, input parameters, form factors, and theoretical uncertainties are discussed in detail.

\subsection{$B_s - \bar B_s$ mixing}
\label{sec:mixing}

\begin{figure}[t]
  \centering
  \subfigure[]{\hspace{-0.8em}\includegraphics[width=0.24\textwidth]{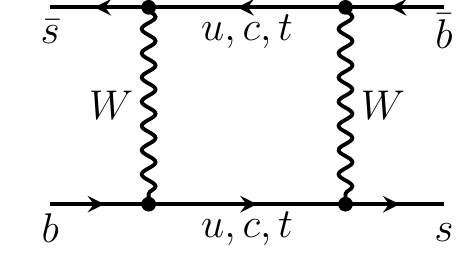}}
  \qquad
  \subfigure[]{\hspace{-0.8em}\includegraphics[width=0.24\textwidth]{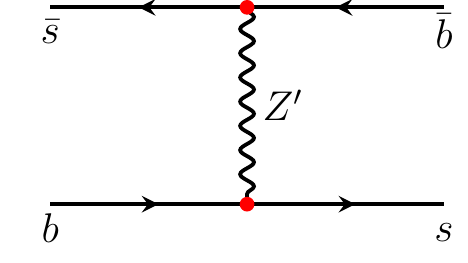}}
  \qquad
  \subfigure[]{\hspace{-0.8em}\includegraphics[width=0.24\textwidth]{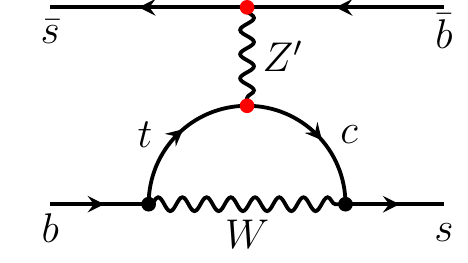}}
  \caption{Feynman diagrams for the $B_s-\bar B_s$ mixing both within the SM (a) and in the $\Zp$ scenario II (b, c). For each diagram, there is also a second one obtained by a $90^\circ$ rotation. }
  \label{fig:Bs-mixing}
\end{figure}

The $B_s - \bar B_s$ mixing is induced by the $W$-box diagram in the SM, and could receive contributions from the tree and penguin diagrams in the $\Zp$ scenario II, as shown in figure~\ref{fig:Bs-mixing}. The effective Hamiltonian for $B_s - \bar B_s$ mixing can be written as~\cite{Buras:2001ra}  
\begin{align}\label{eq:Heff:mixing}
\mathcal H_{\rm eff}^{\Delta B=2} 
= \frac{G_F^2}{16\pi^2}\,m_W^2 (V_{tb}V_{ts}^*)^2\,\mC^{\rm VLL}\, \mO^{\rm VLL} + {\rm h.c.},
\end{align}
with the effective operator $\mO^{\rm VLL}=(\bar s\gamma_\mu P_L b)(\bar s  \gamma^\mu P_L b)$. Analytical expression of the SM Wilson coefficient $\mC_{\rm SM}^{\rm VLL}$ and the QCD renormalization group evolution (RGE) can be found in refs.~\cite{Buchalla:1995vs,Buras:1990fn} and \cite{Buras:2001ra,Ciuchini:1997bw}, respectively. In the $\Zp$ scenario II, the $\Zp$-penguin diagram shown in figure~\ref{fig:Bs-mixing}\,(c) is found to be negligible, and the tree-level $\Zp$-exchange diagram shown in figure~\ref{fig:Bs-mixing}\,(b) results in the NP contribution
\begin{align}
  \mC_{\rm NP,\,II}^{\rm VLL} = \frac{8\pi^2}{G_F^2m_W^2}\frac{1}{(V_{tb}V_{ts}^*)^2}\left(\frac{X_{23}^L}{m_\Zp} \right)^2.
\end{align}
In the $\Zp$ scenario I, the $\Zp$ boson contributes to the $B_s - \bar B_s$ mixing starting at the two-loop level, and its effects are therefore expected to be small. Here we will take $\mC_{\rm NP,\,I} ^{\rm VLL} = 0$.

With the effective Hamiltonian in eq.~(\ref{eq:Heff:mixing}), the off-diagonal mass matrix element of $B_s - \bar B_s$ mixing is given by~\cite{Buras:2001ra}
\begin{align}
  M_{12}^s
= \langle B_s | \mathcal H_{\rm eff}^{\Delta B=2} | \bar B_s \rangle
= \frac{G_F^2}{16\pi^2}\, m_W^2 \big(V_{tb}V_{ts}^*\big)^2\,  \mC^{\rm VLL}\, \langle B_s \left\lvert \mathcal \mO^{\rm VLL} \right\rvert \bar B_s \rangle\,,
\end{align}
where $\mC^{\rm VLL}=\mC_\SM^{\rm VLL}+\mC_\NP^{\rm VLL}$, and the most recent lattice calculations of the hadronic matrix element $\langle B_s \left\lvert \mathcal \mO^{\rm VLL} \right\rvert \bar B_s \rangle$ can be found in ref.~\cite{Aoki:2021kgd}. Then, the mass difference between the two mass eigenstates $B_s^H$ and $B_s^L$ and the CP violation phase read~\cite{Artuso:2015swg}
\begin{align}\label{eq:dltm}
  \Delta m_s = 2 |M_{12}^s|, \qquad \text{and} \qquad \phi_s= \arg M_{12}^s,
\end{align}
respectively. In the case with a complex $\Zp$ coupling $\XctLII$, the phase $\phi_s$ can deviate from the SM prediction and hence affects the CP violation $S_{\psi\phi}$ measured in the decay $B_s \to J/\psi \phi$~\cite{Artuso:2015swg}.

\subsection{$b \to s \nu \bar\nu$ decays}
\label{sec:b2snunu}

\begin{figure}[t]
  \centering
  \subfigure[]{\hspace{-0.8em}\includegraphics[width=0.24\textwidth]{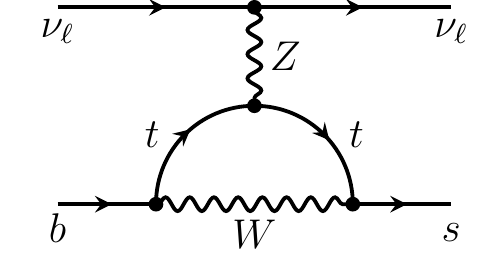}}
  \qquad
  \subfigure[]{\hspace{-0.8em}\includegraphics[width=0.24\textwidth]{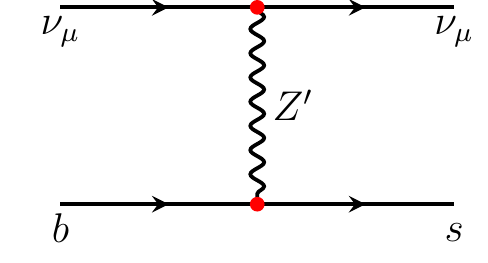}}
  \qquad
  \subfigure[]{\hspace{-0.8em}\includegraphics[width=0.24\textwidth]{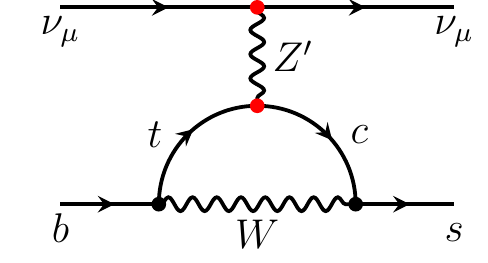}}
  \caption{Feynman diagrams for the $b \to s \nu \bar\nu$ transition, including the selected SM diagram (a), and the $\Zp$ contributions in the scenario II (b, c).}
  \label{fig:bsnunu}
\end{figure}

The rare decays $B \to X_{s} \nu \bar\nu$ and $B \to K^{(*)} \nu \bar\nu$ are all induced by the quark-level $b \to s \nu \bar\nu$ transition. With the $\Zp$ effects taken into account, the effective Hamiltonian governing the $b \to s \nu \bar\nu$ decays can be written as~\cite{Altmannshofer:2009ma,Buras:2014fpa}
\begin{align}\label{eq:Heff:bsnunu}
  \mathcal H_{\rm eff} ^{b \to s \nu \bar\nu} = \frac{4 G_F}{\sqrt 2}\, \frac{\alpha_{\rm e}}{2\pi s_W^2}\, V_{tb}V_{ts}^* \sum_{\ell=e,\mu,\tau} \mC_\ell \big(\bar s \gamma_\mu P_L b\big) \big(\bar \nu_\ell \gamma^\mu P_L \nu_\ell \big) + {\rm h.c.},
\end{align}
where $\ell$ denotes the neutrino flavour. In the SM, the Wilson coefficient $\mC_\ell^\SM(\mu_W)=X(x_t)$ are incuded by the $Z$-penguin and $W$-box diagrams, and are lepton flavour universal. Analytical expression of the Inami-Lim function $X(x_t)$ can be found in refs.~\cite{Buchalla:1998ba,Misiak:1999yg}. Numerically, we obtain $\mC_\ell^\SM(\mu_W)=1.481\pm0.009$~\cite{Buras:2015qea} after including the two-loop electroweak corrections~\cite{Brod:2010hi}. In the $\Zp$ scenario II, the tree-level and the one-loop penguin diagram shown in figure~\ref{fig:bsnunu} contribute to the $b \to s \nu_\mu \bar\nu_\mu$ transition, resulting in 
\begin{align}\label{eq:WC:b2snunu:NP}
  \mC_\mu^{\rm NP,\, II}(\mu_W)= -\frac{1}{\sqrt{2} G_F} \bigg(\frac{1}{8}\frac{V_{cs}^*}{V_{ts}^*}f(x_t) - \frac{\pi s_W^2}{\alpha_{\rm e}V_{tb}V_{ts}^*}\bigg)\frac{X_{23}^L\lambda_{22}^L}{m_{\Zp}^2},
\end{align}
with the loop function $f(x_t)$ defined already by eq.~(\ref{eq:loop}). Numerically, the tree-level $\Zp$-exchange contribution dominates over that from the one-loop $\Zp$-penguin diagram, as in $b \to s \mu^+ \mu^-$ decays. As there are no couplings of the $\Zp$ boson to $\nu_e$ and $\nu_\tau$ neutrinos, $\mC_e^{\NP,\, \rm II}=\mC_\tau^{\NP,\, \rm II}=0$. In the $\Zp$ scenario I, there is no direct coupling of the $\Zp$ boson to neutrinos. The $\Zp$ boson contributes to the $b \to s \nu \bar \nu$ transition at least at the two-loop level and hence can be safely neglected, i.e. $\mC_\ell^{\rm NP,\,I} = 0$.

The inclusive decay $B \to X_s \nu \bar\nu$ is the theoretically cleanest rare $B$-meson decay~\cite{Hurth:2003vb}. With the effective Hamiltonian in eq.~(\ref{eq:Heff:bsnunu}), its differential decay width can be written as~\cite{Altmannshofer:2009ma}
\begin{align}
  \frac{d\Gamma(B\to X_{s} \nu \bar{\nu})}{ds_{b}} =& N(m_b,m_s)\,\kappa(0) \big[3 s_{b}\left(1+\hat{m}_{s}^2-s_{b}\right)+\lambda\left(1, \hat{m}_{s}^2, s_{b}\right)\big]  \sum_{\ell=e,\mu,\tau}\big\lvert \mC_\ell(\mu_W)\big\rvert^2,
\end{align}
with the overall factor given by
\begin{align}
  N(m_1,m_2) = \frac{G_F^2 \alpha_e^2 m_1^5}{384 \pi^5 s_W^4}\big\lvert V_{t b} V_{t s}^*\big\rvert^2 \lambda^{1/2}\bigg(1,\frac{m_2^2}{m_1^2},\frac{q^2}{m_1^2}\bigg),
\end{align}
where $\lambda(x, y, z)=x^2+y^2+z^2-2(x y+y z+z x)$, $\hat{m}_{i}=m_{i} / m_{b}$, and $s_b = q^2/m_b^2$ with $q^2$ the invariant mass squared of the neutrino pair. The factor $\kappa(0)=0.83$ contains the virtual and bremsstrahlung QCD corrections to the $ b \to s\nu\bar{\nu} $ matrix element~\cite{Buchalla:1993bv,Grossman:1995gt}.

For the exclusive decay $B \to K \nu \bar \nu$, the dineutrino invariant mass distribution reads~\cite{Colangelo:1996ay,Altmannshofer:2009ma}
\begin{align}
\frac{d \Gamma(B \to K \nu \bar{\nu})}{d s_B}=\frac{1}{2} N(m_B,m_K)\, \lambda\left(1, \tilde{m}_{K}^2, s_B\right)\left[f_{+}^{B\to K}\left(s_B\right)\right]^2 \sum_{\ell=e,\mu,\tau}\big\lvert \mC_\ell(\mu_W)\big \rvert^2,
\end{align}
where $\tilde{m}_K=m_K / m_B$, $s_B=q^2 / m_B^2$, and $q^2$ denotes the dineutrino invariant mass squared. The $B \to K$ form factor $f_+^{B\to K}(s_B)$ is the main source of theoretical uncertainties. Recent lattice QCD and LCSR results for $f_+^{B\to K}(s_B)$ can be found in refs.~\cite{Bouchard:2013eph,Bailey:2015dka} and \cite{Khodjamirian:2017fxg,Gubernari:2018wyi}, respectively.

For the $B \to K^* \nu \bar{\nu}$ decay, the dineutrino invariant mass spectrum can be written as~\cite{Altmannshofer:2009ma}
\begin{align}
\frac{d \Gamma(B \to K^* \nu \bar\nu)}{d s_B}=\frac{1}{8} N (m_B,m_{K^*})\, s_B \sum_{\ell=e,\mu,\tau}\Big(\left|A^\ell_{\perp}(s_B)\right|^2+\left|A^\ell_{\|}(s_B)\right|^2+\left|A^\ell_0(s_B)\right|^2\Big),
\end{align}
with the three transversity amplitudes given, respectively, as
\begin{align}
A_\perp(s_B) &=-2 \sqrt2\, \lambda^{1 / 2}(1, \tilde{m}_{K^{*}}^2, s_B)\, \mC_\ell(\mu_W) \frac{V(s_B)}{1+\tilde{m}_{K^*}}, \\
A_{\|}(s_B) &=2 \sqrt2\,(1+\tilde{m}_{K^*})\, \mC_\ell(\mu_W)\, A_1(s_B), \nonumber \\
A_{0}(s_B) &=\frac{ \mC_\ell(\mu_W)}{\widetilde{m}_{K^{*}} \sqrt{s_B}}\left[(1-\tilde{m}_{K^{*}}^2-s_B)(1+\tilde{m}_{K^{*}}) A_{1}(s_B)-\lambda(1, \tilde{m}_{K^{*}}^2, s_B) \frac{A_2(s_B)}{1+\tilde{m}_{K^{*}}}\right], \nonumber
\end{align}
where $\tilde{m}_{K^*}=m_{K^*}/m_B$. Here, the main theoretical uncertainties arise from the three $B \to K^*$ transition form factors $V(s_B), A_1(s_B)$ and $A_2(s_B)$. Recent lattice QCD and LCSR calculations of these form factors can be found in refs.~\cite{Horgan:2013hoa,Aoki:2021kgd} and refs.~\cite{Bharucha:2015bzk,Gubernari:2018wyi,Gao:2019lta}, respectively.

\subsection{$s \to d \mu^+ \mu^-$ transition}

\begin{figure}[t]
  \centering
  \subfigure[]{\hspace{-0.8em}\includegraphics[width=0.24\textwidth]{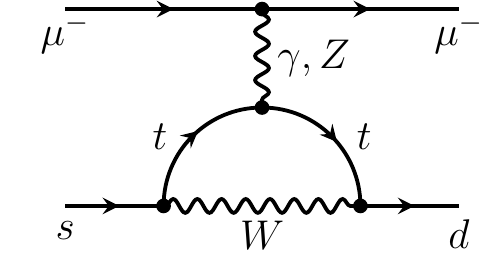}}
  \quad
  \subfigure[]{\hspace{-0.8em}\includegraphics[width=0.24\textwidth]{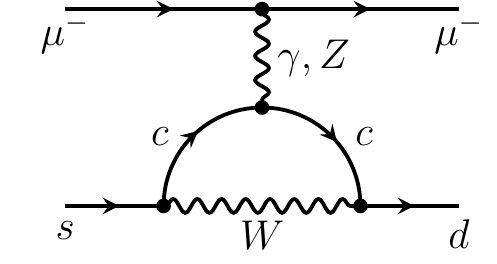}}
  \quad
  \subfigure[]{\hspace{-0.8em}\includegraphics[width=0.24\textwidth]{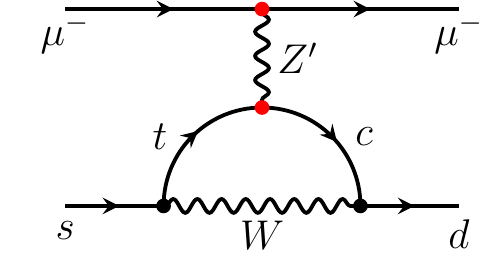}}
  \quad
  \subfigure[]{\hspace{-0.8em}\includegraphics[width=0.24\textwidth]{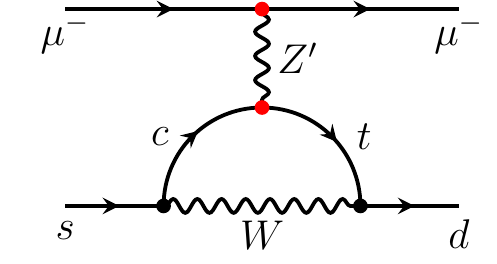}}
  \caption{Feynman diagrams for the $s \to d \mu^+ \mu^-$ transition, including the selected SM diagrams (a, b), and the $\Zp$ contributions (c, d).}
  \label{fig:sdmumu}
\end{figure}

For the $s \to d \mu^+ \mu^-$ transition, the most relevant decay modes are the $ K_{L,S} \to \mu^+ \mu^- $ decays. After taking into account the $\Zp$ effects, the effective Hamiltonian inducing the short-distance (SD) contribution to $K_{L,S}\to \mu^+ \mu^- $ decays can be written as~\cite{Buchalla:1993wq}
\begin{align}\label{eq:Heff:sdll}
  \mathcal H_{\rm eff}^{s \to d \mu^+ \mu^-} =\frac{4 G_F}{\sqrt 2} \frac{\alpha_{\rm e}}{4\pi s_W^2} \, \big(V_{ts}^*V_{td}\,Y+V_{cs}^*V_{cd}\,Y_{\rm NL} \big)\big(\bar s \gamma_\mu P_L d\big) \big(\bar \mu \gamma^\mu \gamma_5 \mu \big)  + {\rm h.c.}.
\end{align}
In the SM, the function $Y=Y(x_t)$ describes contributions from the penguin diagrams with internal top quark~\cite{Misiak:1999yg,Buchalla:1998ba,Hermann:2013kca,Bobeth:2013tba}, while $Y_{\rm NL}$ involves the charm-quark contributions~\cite{Gorbahn:2006bm}. In both the $\Zp$ scenarios I and II, the $s \to d \mu^+ \mu^-$ transition is induced by the penguin diagrams shown in figure~\ref{fig:sdmumu}, which result in the NP contributions
\begin{align}\label{eq:Y:NP}
  Y_{\rm NP,I} &= \frac{1}{8\sqrt2G_F }\bigg(\frac{V_{cd}}{V_{td}}\frac{\XctL^* \lmmL}{m_\Zp^2}+\frac{V_{cs}^*}{V_{ts}^*}\frac{\XctL \lmmL}{m_{\Zp}^2}\bigg)f(x_t),  \\
  Y_{\rm NP,II}&=\frac{1}{8\sqrt2G_F }\bigg(\frac{V_{cd}}{V_{td}}\frac{\XctLII^* \lmmLII}{m_{\Zp}^2}+\frac{V_{cs}^*}{V_{ts}^*}\frac{\XctLII \lmmLII}{m_{\Zp}^2}\bigg)f(x_t),
\end{align}
with the loop function $f(x_t)$ defined already in eq.~(\ref{eq:loop}). Here, the NP contributions in both of the two $\Zp$ scenarios are of the same magnitude, which is different from what is observed in the $b \to s \mu^+ \mu^-$ transitions. Numerically, $Y=0.94-(1.03+0.02i)\XctL\lmmL-(0.99+0.45i)\XctL^*\lmmL$ for $m_\Zp=1\TeV$ in scenario I.

For the $K_{L,S} \to \mu^+ \mu^- $ decays, only the SD part of a dispersive contribution can be reliably calculated. The branching ratio of $K_L \to \mu^+ \mu^-$ can be written as~\cite{Buchalla:1993wq}
\begin{align}
\mB(K_L \to \mu^+ \mu^-)_{\rm SD}=\kappa_\mu\left[\frac{{\rm Re}(\lambda_t Y)}{\lambda^5}+\frac{{\rm Re} \lambda_c}{\lambda} P_c(Y)\right]^2,
\end{align}
with $\lambda_t=V_{ts}^*V_{td}$, $\lambda_c=V_{cs}^*V_{cd}$, and $\lambda \approx V_{us}$ denoting the Wolfenstein parameter. The factor $\kappa_\mu$ contains the relevant hadronic matrix element that can be extracted from the $K^+ \to \mu^+ \nu_\mu$ decay, and numerically we have $\kappa_\mu=(2.009 \pm 0.017) \times 10^{-9}\, (\lambda/0.225)^8$~\cite{Buchalla:1993wq,Gorbahn:2006bm}.
The charm contribution $P_c(Y)$ is found to be $P_c(Y) =0.115 \pm 0.017$ at the NNLO in QCD~\cite{Gorbahn:2006bm}. For the $ K_S \to \mu^+ \mu^- $ decay, the SD and long-distance contributions add incoherently in the total rate~\cite{Isidori:2003ts,Ecker:1991ru,DAmbrosio:2017klp,Dery:2021mct}. The SD part of the branching ratio is given by~\cite{Isidori:2003ts,Ecker:1991ru,DAmbrosio:2017klp,Dery:2021mct}
\begin{align}
\mB (K_S \to \mu^+ \mu ^- )_{\rm SD}=\tau_{K_S} \frac{G_F^2 \alpha_{\rm e}^2}{8 \pi^3s_W^4} m_K f_K^2 \bigg(1-\frac{m_\mu^2}{m_K^2}\bigg)^{1/2} m_\mu^2 \, {\rm Im}^2\big(\lambda_t Y\big),
\end{align}
with $\tau_{K_S}$ the lifetime of $K_S$ and $f_K $ the decay constant.

\subsection{$s \to d \nu \bar\nu$ decays}

The $K^+ \to \pi^+ \nu \bar\nu $ and $K_L \to \pi^0 \nu \bar\nu$ decays are induced by the $s \to d \nu\bar\nu$ transition. They are both characterized by the theoretically clean virtue, since the relevant hadronic matrix elements can be extracted with the help of isospin symmetry from the leading semi-leptonic $K_{\ell 3}$ decays~\cite{Buras:2004uu}. With the $\Zp$ effects taken into account, the $s \to d \nu \bar\nu$ decays are governed by the effective Hamiltonian~\cite{Buchalla:1998ba,Misiak:1999yg}
\begin{align}\label{eq:Heff:sdnunu}
  \mathcal H_{\rm eff}^{s \to d \nu \bar\nu}=\frac{4 G_F}{\sqrt 2} \frac{\alpha_{\rm e}}{2\pi s_W^2} \sum_{\ell =e, \mu, \tau}\big( V_{ts}^* V_{td} X^\ell+V_{cs}^*V_{cd}X_{\rm NL}^\ell\big) \big( \bar s \gamma_\mu P_L d \big) \big( \bar\nu_\ell \gamma^\mu P_L \nu_\ell \big) + {\rm h.c.}.
\end{align}
In the SM, similar to the $b \to s \nu \bar\nu$ transition, the $Z$-penguin and $W$-box diagrams with internal top quark result in a flavour-universal Wilson coefficient $X_{\rm SM}^\ell=X(x_t)$, with the Inami-Lim function $X(x_t)$ introduced already in eq.~(\ref{eq:Heff:bsnunu}). The contribution from internal charm quark is represented by the function $X_{\rm NL}^\ell$.  In the $\Zp$ scenario II, the $\Zp$ boson can only interact with the $\nu_\mu$ neutrino. Its contributions to the operators with $\nu_e$ and $\nu_\tau$ neutrinos arise from two-loop level and can be therefore neglected, i.e., $X_{\NP, \rm II}^e=X_{\NP,\rm II}^\tau=0$. However, the $\Zp$-penguin diagrams similar to the one shown in figure~\ref{fig:sdmumu} contribute to the Wilson coefficient $X^\mu$, and we find $X_{\rm NP,II}^\mu=-Y_{\rm NP}^{\rm II}$. In the $\Zp$ scenario I, as in the case of the $b \to s \nu \bar\nu$ transition, the $\Zp$ effects arise firstly at two-loop level and are neglected, i.e., $X_{\rm NP, I}^\ell=0$.

In the SM, with the help of isospin symmetry, the branching ratio of $K^+ \to \pi^+ \nu \bar\nu$ decay after summing over the three neutrino flavours is given by~\cite{Buras:2004uu,Mescia:2007kn,Cirigliano:2011ny} 
\begin{equation}
\mB (K^+ \to \pi^+ \nu \bar\nu )=\frac{1}{3}\kappa_+(1+\Delta_{\rm EM}) \sum_{\ell=e,\mu,\tau} \bigg[\bigg(\frac{{\rm Im} [\lambda_t X^\ell]}{\lambda^5} \bigg)^2+ \bigg(\frac{{\rm Re} \lambda_c}{\lambda} P_c(X)+\frac{{\rm Re} [\lambda_tX^\ell]}{\lambda^5}\bigg)^2\,\bigg].
\end{equation}
The parameter $\kappa_+$ contains the relevant hadronic matrix elements extracted from the $K_{\ell 3}$ decays, and numerically we have $\kappa_+=(5.173\pm 0.025)\times 10^{-11}\,(\lambda/0.225)^8$~\cite{Buras:2015qea}. The parameter $\Delta_{\rm EM}=-0.003$ describes the isospin breaking correction~\cite{Mescia:2007kn,Cirigliano:2011ny}. The parameter $P_c(X)$ summarizes the SD charm contribution~\cite{Buchalla:1998ba,Buchalla:1993wq,Buras:2006gb,Brod:2008ss} and the long-distance contributions~\cite{Isidori:2005xm}. Numerically, it is found that $P_c(X)=0.404 \pm 0.024$~\cite{Buras:2015qea}.

The rare decay $K_L \to \pi^0 \nu \bar\nu$ proceeds in the SM almost entirely through direct CP violation~\cite{Littenberg:1989ix,Buchalla:1998ux}. It is completely dominated by the SD loop diagrams with top-quark exchanges, and the charm contribution can be fully neglected~\cite{Buras:2004uu}. With the help of isospin symmetry, the branching ratio of $K_L \to \pi^0 \nu \bar\nu$ decay after summing over three neutrino flavours reads~\cite{Buras:2004uu,Buchalla:1996fp}
\begin{align}
\mB (K_L \to \pi^0 \nu \bar\nu )=\frac{1}{3}\kappa_L (1-\delta_\epsilon) \sum_{\ell=e,\mu,\tau}\left(\frac{1}{\lambda^5} \text{Im} \big[\lambda_t X^\ell\big]\right)^2,
\end{align}
where $\kappa_L$ encodes the hadronic matrix element extracted from the $K_{\ell3} $ data~\cite{Mescia:2007kn,Marciano:1996wy}, and numerically $\kappa_{L}=(2.231 \pm 0.013) \times 10^{-10}\, (\lambda/0.225)^8$. The parameter $\delta_\epsilon$ denotes the indirectly CP-violating contribution, and is highly suppressed by the $K^{0}-\bar{K}^{0}$ mixing parameter $ |\epsilon| $~\cite{Buchalla:1998ux}.

\subsection{$t \to c \mu^+ \mu^-$ decay}

\begin{figure}[t]
  \centering
  \includegraphics[width=0.25\textwidth]{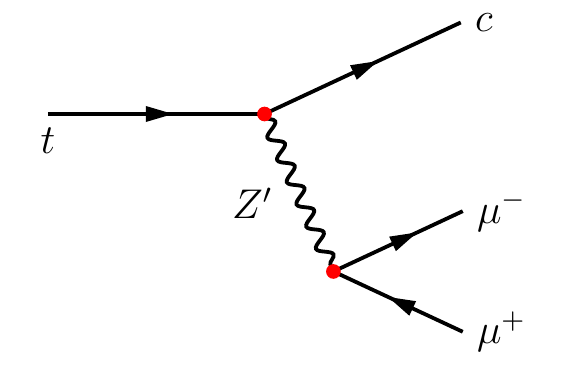}
  \qquad\qquad
  \includegraphics[width=0.25\textwidth]{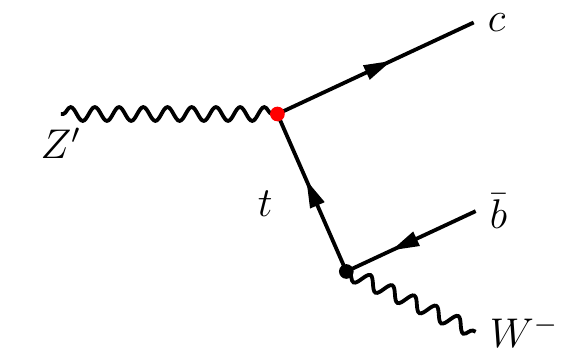}
  \caption{Feynman diagrams for the $t \to c \mu^+ \mu^-$ (left) and $\Zp \to \bar b c W^-$ (right) decays in the $\Zp$ scenarios.}
  \label{fig:top}
\end{figure}

In the SM, the rare FCNC decay $t \to c \mu^+ \mu^-$ is highly suppressed by the Glashow-Iliopoulos-Maiani mechanism~\cite{Glashow:1970gm}, with a branching ratio of $\mathcal{O}(10^{-10})$~\cite{Eilam:1990zc,Aguilar-Saavedra:2004mfd}. However, this process could be significantly enhanced by the $\Zp$ boson through the tree-level diagrams shown in figure~\ref{fig:top}. Let us discuss the $\Zp$ effects in the cases $m_\Zp > m_t$ and $m_\Zp < m_t$, respectively.

\subsubsection{$m_\Zp > m_t$}

The branching ratio of the $\Zp$-mediated $t \to c \mu^+ \mu^-$ decay can be written as
\begin{align}
 \mB(t \to c \mu^+ \mu^-) = \frac{\Gamma(t \to c \mu^+ \mu^-)}{\Gamma_t},
\end{align}
where $\Gamma_t$ is the total width of the top quark. The differential decay width can be calculated from the left tree-level diagram shown in figure~\ref{fig:top}, with the result in the $\Zp$ scenario I given by
\begin{align}\label{eq:width:LO}
\frac{d \Gamma_0^{\rm I}(t \to c \mu^+ \mu^-) }{d q^2}&=\frac{\left|\XctL \lambda^L_{\mu\mu}\right| ^2}{768\pi^{3}m_t^3}\frac{(m_t^2-q^2)(m_t^4+q^2m_t^2-2q^4)}{(q^2-m_{\Zp}^2)^2+m_\Zp^2 \Gamma_\Zp^2},
\end{align}
where $q^2$ denotes the dilepton invariant mass squared, and $\Gamma_{\Zp}$ the finite decay width of the $\Zp$ boson. Since the $\Zp$ boson cannot be on-shell in the case of $m_\Zp > m_t$, we can safely neglect the finite-width effect of the $\Zp$ boson here. After including the NLO QCD correction, the differential decay width of $t \to c \mu^+ \mu^-$ can be rewritten as
\begin{align}\label{eq:width:NLO}
  \frac{d\Gamma_{\rm NLO}^{\rm I}(t \to c \mu^+ \mu^-)}{d q^2} = \frac{d \Gamma_0^{\rm I}(t \to c \mu^+ \mu^-) }{d q^2} \big[ 1+f_{\rm NLO}(q^2) \big],
\end{align}
where the factor $f_{\rm NLO}(q^2)$ can be obtained from the NLO QCD calculation of the $t \to c Z$ decay~\cite{Zhang:2010bm,Drobnak:2010wh,Drobnak:2010by}, and is given explicitly by
\begin{align}\label{eq:NLO correction}
  f_{\rm NLO}(q^2) =  \frac{4\alpha_{s}}{3 \pi}\biggl[&-\frac{1}{2} \pi^2-3 -\frac{1}{2}\log \frac{\mu^2}{m_t^2} +\frac{12+135 \beta^2-43 \beta^4}{12 \beta^2 (3-\beta^2)} +\left(2 \log \beta-\frac{9-\beta^2}{3-\beta^2}\right) \log \beta \nonumber \\
  &  +\frac{(1-\beta^2)(1-4\beta^2+\beta^4)}{\beta^4\left(3-\beta^2\right)} \log (1-\beta^2) + \mathrm{Li}_2(\beta^2)+ \mathrm{Li}_2(1- \beta^{-2})         \biggr],
\end{align}
with $ \beta= \sqrt{1-q^2/m_t^2}$. Numerically, the NLO QCD corrections decrease the LO width by $6.4\%\sim 9.1\%$ for $m_t < m_\Zp < 1\TeV$. Expressions in scenario II can be obtained from the above formulas with the replacement $(\XctL, \lmmL) \to (X_{23}^L,\lambda_{22}^L)$. As will be shown in subsection~\ref{sec:num:high energy:heavy Zp}, the branching ratio $\mB(t \to c \mu^+ \mu^-)$ is predicted to be below $\mO(10^{-5})$ in the two $\Zp$ scenarios. Therefore, their contributions to the top-quark total width $\Gamma_t$ can be safely neglected.

\subsubsection{$m_\Zp < m_t$}
\label{sec:t2cmumu:light Zp}

In the case of $m_\Zp < m_t$, the branching ratio of the $\Zp$-mediated $t \to c \mu^+ \mu^- $ decay can also be written as eqs.~(\ref{eq:width:LO}) and (\ref{eq:width:NLO}). However, since the intermediate $\Zp$ boson can be on-shell in this case, it is necessary to consider the $\Zp$ finite-width effect in eq.~(\ref{eq:width:LO}). When $m_W < m_\Zp < m_t$, the main $\Zp$ decay modes in scenario I are $\Zp \to \mu^+ \mu^-$ and $\Zp \to  b \bar c W^+$ decays. Therefore, the total width of the $\Zp$ boson can be written as
\begin{align}
  \Gamma_\Zp^{\rm I} = \Gamma^{\rm I}(\Zp \to \mu^+ \mu^-) + 2 \Gamma^{\rm I}(\Zp \to b \bar c W^+),
\end{align}
where the decay width of $\Zp \to \mu^+ \mu^-$ is given by
\begin{align}\label{eq:Zp to mumu:I}
  \Gamma^{\rm I} (\Zp \to \mu^+ \mu^-) &= \frac{1}{24\pi} \left| \lambda^L_{\mu\mu}\right| ^2m_\Zp,
\end{align}
and the width of the top-quark mediated $\Zp \to b \bar c W^+$ decay reads
\begin{align}\label{eq:Zp to bcW:I}
  \Gamma^{\rm I}(\Zp \to b \bar c W^+)=
  &\frac{G_F |V_{tb}|^2|X_{ct}|^2}{128\sqrt{2}\,\pi^3}\frac{m_t^8}{m_\Zp^5}\biggl\lbrace 4 y_W^3y_\Zp^3 \log \frac{y_\Zp}{y_W}  +\frac{1}{12}(y_\Zp-y_W)\Big[ 60 + 48y_W^2 y_\Zp^2 \nonumber
  \\
  &  +20y_W y_\Zp+(30-39y_Wy_\Zp)(y_W+y_\Zp)-88(y_W^2+y_\Zp^2) +9(y_W^3+y_\Zp^3)\Big]
  \nonumber\\
  & +\Big[ 5+ 9 y_W^2y_\Zp^2 -9 \big(y_W^2+y_\Zp^2\big)-4y_W^3y_\Zp^3+4 \big( y_W^3+y_\Zp^3 \big)\Big]\log\frac{\beta_\Zp^2}{\beta_W^2}\biggr\rbrace,
\end{align}
with $\beta_W=\sqrt{1-m_W^2/m_t^2}$, $\beta_\Zp=\sqrt{1-m_\Zp^2/m_t^2}$, $y_W=m_W^2/m_t^2$, and $y_\Zp=m_\Zp^2/m_t^2$.

In scenario II, the $\Zp$ boson can also decay into the additional channels $\Zp \to \nu_\mu \bar\nu_\mu$ and $\Zp \to b \bar s$. The total width can then be written as
\begin{align}
\Gamma_\Zp^{\rm II} = \Gamma^{\rm II}(\Zp \to \mu^+ \mu^-) +\Gamma^{\rm II}(\Zp \to \nu_\mu \bar\nu_\mu) + 2 \Gamma^{\rm II}(\Zp \to \bar c b W^+)+2\Gamma^{\rm II} (\Zp \to b \bar s),
\end{align}
with 
\begin{align}
  \Gamma^{\rm II} (\Zp \to \nu_\mu \bar\nu_\mu) = \frac{1 }{24\pi} \left| \lambda^L_{22}\right|^2 m_\Zp
  \qquad \text{and}, \qquad
  \Gamma^{\rm II} (\Zp \to b \bar s) = \frac{1}{8\pi}  \left|X_{23}^L\right| ^2 m_\Zp.
\end{align}
The decay widths $\Gamma^{\rm II}(\Zp \to \mu^+ \mu^-)$ and $\Gamma^{\rm II}(\Zp \to \bar c b W^+)$ can be obtained from eqs.~(\ref{eq:Zp to mumu:I}) and (\ref{eq:Zp to bcW:I}) with the replacement $ (\lmmL, \XctL) \to (\lmmLII, \XctLII)$.

In the SM, $t\to bW$ is the main decay channel of the top quark. When $m_\Zp < m_t$, the top quark can also decay into an on-shell $\Zp$, which contributes to the top-quark total width as 
\begin{align}
  \Gamma_t \simeq \Gamma(t \to b W) + \Gamma( t \to c \Zp).
\end{align}
After including the NLO QCD correction, we can write the decay width of $t \to b W$ as~\cite{Jezabek:1988iv,Czarnecki:1990kv,Li:1990qf}
\begin{align}
  \Gamma (t\to b W) =&\Gamma_0(t\to b W)\Bigg\lbrace 1+\dfrac{2\alpha_{s}}{3\pi}\left[ 2\left( \dfrac{(1-\beta_W^2)(2\beta_W^2-1)(\beta_W^2-2)}{\beta_W^4(3-2\beta_W^2)}\right)\log(1-\beta_W^2)\right.  \nonumber\\
                     &\left.-\dfrac{9-4\beta_W^2}{3-2\beta_W^2}\log\beta_W^2+2\text{Li}_2\beta_W^2-2\text{Li}_2(1-\beta_W^2)-\dfrac{6\beta_W^4-3\beta_W^2-8}{2\beta_W^2(3-2\beta_W^2)}-\pi^2 \right] \Bigg\rbrace,
\end{align}
with the LO result given by
\begin{align}
\Gamma_0 (t\to b W) = \frac{G_F m^3_t}{8\sqrt{2}\,\pi} \vert V_{tb}\vert^2 \beta^4_W (3-2\beta^2_W).
\end{align}
For the $t \to c \Zp$ decay in scenario I, the decay width including the NLO QCD correction can be written as
\begin{align}
  \Gamma_{\rm NLO}^{\rm I}(t \to c \Zp) = \Gamma_0^{\rm I}(t \to c \Zp) \big[1+f_{\rm NLO}(m_\Zp^2)\big],
\end{align}
with the tree-level result given by
\begin{align}
  \Gamma_0^{\rm I} (t \to c \Zp) = \frac{m_t^3}{32\pi m_\Zp^2}| X_{ct}|^2 \beta_\Zp^4 \big(3-2\beta_\Zp^2 \big).
\end{align}
The factor $f_{\rm NLO}(q^2)$ is obtained from the calculation of NLO QCD correction to the $t \to c Z$ decay~\cite{Zhang:2010bm,Drobnak:2010wh,Drobnak:2010by} and its explicit expression has been given in eq.~(\ref{eq:NLO correction}). Numerically, we find $f(m_\Zp^2) \approx -10.7\% \sim + 5.3\%$ when $m_\Zp$ varies in the range $105\GeV < m_\Zp < m_t$. Expressions in the $\Zp$ scenario II can be obtained from the above formulas with the replacement $(\XctL, \lmmL) \to (\XctLII,\lmmLII)$. Since the decay rate of $t \to c \Zp$ is proportional to $|X_{ct}^L|^2$ in scenario I, the top-quark width can provide a unique constraint on the $t\bar{c}\Zp$ coupling in the case of $m_\Zp < m_t$.

\section{Numerical results and discussions}
\label{sec:num}

\begin{table}[t]
\tabcolsep 0.15in
\renewcommand*{\arraystretch}{1.3}
\centering
\begin{tabular}{c c c c}
  \hline\hline
  Input & Value & Unit & Ref.
  \\\hline
  $m_t^{\text{pole}}$ & $173.76 \pm 0.30$& GeV & \cite{PDG:2020}
  \\
  $m_b(m_b)$& $4.195 \pm 0.014$  & GeV & \cite{FermilabLattice:2018est} 
  \\
  $m_c(m_c)$ & $1.273 \pm 0.01$ & GeV &  \cite{FermilabLattice:2018est} 
  \\\hline
  $|V_{cb}|$(semi-leptonic) & $(41.1 \pm 0.3 \pm 0.5) \times 10^{-3}$& &  \cite{Charles:2004jd}
  \\
 $|V_{ub}|$(semi-leptonic) & $(3.91 \pm 0.08 \pm 0.21) \times 10^{-3} $ & &  \cite{Charles:2004jd}
  \\
  $|V_{us}|f^{K \to \pi}_+$(0) & $0.2165 \pm 0.0004 $ & & \cite{Charles:2004jd}
  \\
  $\gamma$&  $72.1^{+5.4}_{-5.7} $ & ${}^\circ$ & \cite{Charles:2004jd}
  \\
  $f^{K \to \pi}_+(0)$  & $0.9681 \pm 0.0014 \pm 0.0022$ &  & \cite{Charles:2004jd}
  \\\hline\hline
\end{tabular}
\caption{Main input parameters used in our numerical analysis.}
\label{tab:input}
\end{table}

In this section, we proceed to present our numerical analysis of the $\Zp$ scenarios presented in section~\ref{sec:Zp}. In table~\ref{tab:input}, we list the main input parameters used throughout this work. Table~\ref{table:observables} summaries the SM predictions and the current experimental data for the various processes discussed in the previous section.

\begin{table}[t]
  \tabcolsep 0.15in
  \renewcommand*{\arraystretch}{1.3}
  \centering
  \begin{tabular}{c|c|c|c}
    \hline\hline
    Observable    &  SM & Exp. & Ref. \\
    \hline
    $\mathcal B (B_s\to\mu^+\mu^-)$ & $(3.46 \pm 0.11) \times 10^{-9}$ & $(3.09^{+0.46 +0.15}_{-0.43 -0.11})\times 10^{-9}$ & \cite{LHCb:2021vsc}
    \\ 
    \hline
    $\Delta m_s $ [GeV] &$(1.18 \pm 0.07)\times 10^{-11}$  & $(1.1688 \pm 0.0014)\times 10^{-11}$ & \cite{Amhis:2019ckw}\\
    $S_{\psi\phi}$ & $0.041 \pm 0.003$ & $0.033 \pm 0.033 $ & \cite{Amhis:2019ckw} \\
    \hline
    $\mathcal{B}(B \to X_s \nu \bar{\nu}) $  & $(3.03 \pm 0.08)\times 10^{-5\hphantom{1}}$  & $<6.4\times 10^{-4}$ & \cite{Barate:2000rc} \\
    $\mathcal{B}(B^+ \to K^+ \nu \bar{\nu}) $  & $(4.15 \pm 0.56)\times 10^{-6\hphantom{1}}$  & $<1.9 \times 10^{-5}$ &\cite{Belle:2017oht}    \\
    $\mathcal{B}(B^0 \to K^0 \nu \bar{\nu}) $ & $(3.85 \pm 0.52)\times 10^{-6\hphantom{1}}$  &$<2.6 \times 10^{-5}$ & \cite{Belle:2017oht}   \\
    $\mathcal{B}(B^+ \to K^{*+} \nu \bar{\nu}) $ & $(9.71 \pm 0.93) \times 10^{-6\hphantom{1}}$ & $<6.1 \times 10^{-5}$ & \cite{Belle:2017oht} \\
    $\mathcal{B}(B^0 \to K^{*0} \nu \bar{\nu}) $ & $(9.01 \pm 0.87) \times 10^{-6\hphantom{1}}$  &$<1.8 \times 10^{-5}$ & \cite{Belle:2017oht}\\
    \hline
    $\mathcal{B}(K_L \to  \mu^+ \mu^-) $  &$(7.34 \pm 1.21) \times 10^{-9\hphantom{1}}$  & $(6.84 \pm 0.11)\times 10^{-9}$ & \cite{PDG:2020} \\
    $\mathcal{B}(K_S \to  \mu^+ \mu^-) $   & $(5.18 \pm 1.57)\times 10^{-12}$  & $(0.94^{+0.72}_{-0.64})\times 10^{-10}$ & \cite{LHCb:2020ycd}   \\
    \hline
    $\mathcal{B}(K^+ \to  \pi^+ \nu \bar{\nu}) $   &$(8.38 \pm 0.62)\times 10^{-11}$   & $(10.6 ^{+4.0}_{-3.4} \pm0.9)\times 10^{-11}$ & \cite{CortinaGil:2021nts}   \\
    $\mathcal{B}(K_L \to  \pi^0 \nu \bar{\nu}) $   &$(3.46 \pm 0.46)\times 10^{-11}$   & $<3.0\times 10^{-9}$ & \cite{Ahn:2018mvc} 
    \\\hline\hline
  \end{tabular}
  \caption{Our SM predictions and the experimental measurements of the relevant observables used in our global fit. Upper limits are all at $95\%$ confidence level (CL) .}
  \label{table:observables}
\end{table}

As discussed in section~\ref{sec:Zp}, the relevant model parameters in $\Zp$ scenario I (II) contain the complex couplings $\XctL$ and $\lmmL$ ($X_{23}^L$ and $\lambda_{22}^L$), and the $\Zp$ mass $m_\Zp$. When performing a global fit on these parameters and studying their effects, we consider the following parameter space:
\begin{align}\label{eq:parameter space}
  \text{S.I: }\qquad &\lvert \XctL \rvert < 2.0, \qquad \lvert\lmmL \rvert < 2.0, \qquad 105\GeV < m_\Zp < 1000\GeV,
  \\\nonumber
  \text{S.II: }\qquad & \lvert \XctLII \rvert < 2.0, \qquad \lvert \lmmLII \rvert < 2.0, \qquad 105\GeV < m_\Zp < 1000\GeV,
\end{align}
where the upper limits on the magnitudes of the couplings correspond to about half of the perturbativity bound $\sqrt{4\pi}$. Here the lower bound of $m_\Zp$ is chosen to avoid the potentially strong bound from the branching ratio $\mB(t \to c Z)$~\cite{ATLAS:2021stq} (see subsection~\ref{sec:num:high energy:heavy Zp} for further details).

In our numerical analysis, we follow the approach used in ref.~\cite{Altmannshofer:2014rta} and construct a likelihood function that only depends on the Wilson coefficients as
\begin{align}\label{eq:likelihood}
  -2\log L(\boldsymbol{\mC}) = \sum_i \boldsymbol{x_i}^T (\boldsymbol{\mC}) \big[ V_{\rm exp} + V_{\rm th}(\boldsymbol{\theta})]^{-1} \boldsymbol{x_i}(\boldsymbol{\mC}),
\end{align}
with the definition
\begin{align*}
  \boldsymbol{x}_i(\boldsymbol{\mC})=\boldsymbol{\mO}_i^{\rm exp} - \boldsymbol{\mO}_i^{\rm th}(\boldsymbol{\mC}, \boldsymbol{\theta}).
\end{align*}
Here, $\boldsymbol{\mO}_i^{\rm exp}$ are the central values of the experimental measurements, while  $\boldsymbol{\mO}_i^{\rm th}$ denote the central values of the theoretical predictions, which depend on the Wilson coefficients $\boldsymbol{\mC}$ and the input parameters $\boldsymbol{\theta}$. $V_{\rm exp}$ is the covariance matrix of the experimental measurements, while $V_{\rm th}$ denotes the covariance matrix of the theoretical predictions, which contains all the theoretical uncertainties and their correlations. Generally, the covariance matrix $V_{\rm th}$ depends on both the Wilson coefficients $\boldsymbol{\mC}$ and the input parameters $\boldsymbol{\theta}$. However, as all the experimental measurements do not show large deviations from the SM predictions, we have neglected in our numerical analysis the dependence of $V_{\rm th}$ on the $\Zp$ contributions to the Wilson coefficients $\boldsymbol{\mC}$, i.e., $V_{\rm th}$ is evaluated with the Wilson coefficients $\boldsymbol{\mC}$ fixed to their SM values. Furthermore, the theoretical uncertainties are approximated as Gaussian and obtained by randomly sampling the observables with the input parameters $\boldsymbol{\theta}$ distributed according to their probability density functions. With eq.~(\ref{eq:likelihood}), the $\Delta\chi^2$ function, which depends on the Wilson coefficients $\boldsymbol{\mC}$, can be written as $\Delta\chi^2(\boldsymbol{\mC})=-2\log L(\boldsymbol{\mC})/L_{\rm max}$, where $L_{\rm max}$ denotes the maximum of $L$ at different values of the Wilson coefficients. For more details on the fit methodology, we refer to refs.~\cite{Altmannshofer:2014rta,Straub:2018kue}. Our global fit is performed using an extended version of the package \flavio~\cite{Straub:2018kue}.

\subsection{Flavour constraints}
\label{sec:num:low energy}

As discussed in section~\ref{sec:framework}, the $\Zp$ contributions to the $b \to s \mu^+ \mu^-$, $s \to d \mu^+ \mu^-$, $b \to s \nu \bar\nu$, and $s \to d \nu \bar\nu$ transitions are all controlled by the products $\XctL\lmmL/m_\Zp^2$ in scenario I or $\XctLII \lmmLII/m_\Zp^2$ in scenario II. These two products are complex in general and constraints on them will be derived in terms of their real and imaginary parts in the following analysis.

\begin{figure}[t]
	\centering
	\subfigure{\includegraphics[width=0.45\textwidth]{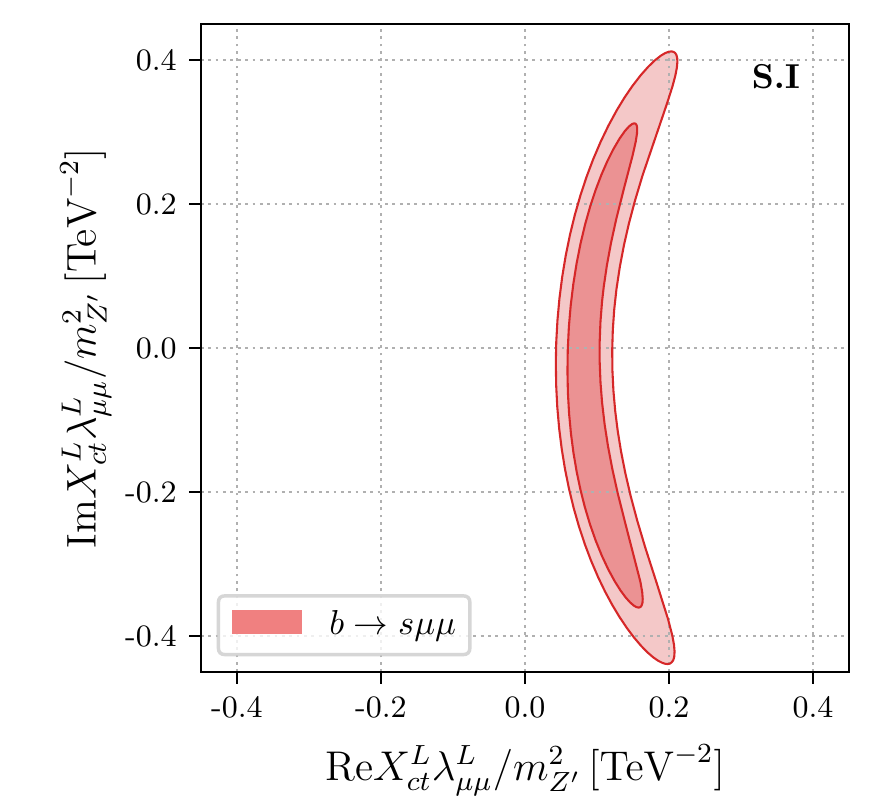}}
	\qquad
	\subfigure{\includegraphics[width=0.45\textwidth]{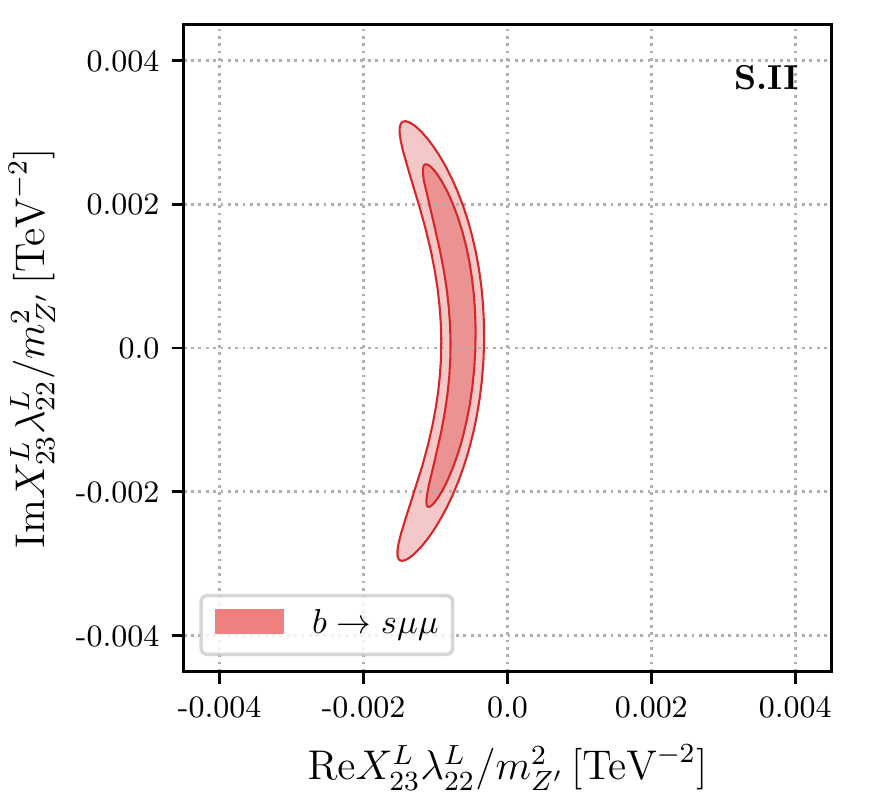}}
	\caption{Allowed regions of the real and imaginary parts of $\XctL \lmmL/m_\Zp^2$ and $X_{23}^L\lambda^{L}_{22}/m_\Zp^2$ by the $b \to s \mu^+ \mu^-$ processes in the $\Zp$ scenarios I (left) and II (right). The dark and the light regions correspond to the $68\%$ and the $95\%$ CL, respectively.} 
	\label{fig:bound:b2smumu}
\end{figure}

For the $b \to s \mu^+ \mu^-$ transitions, as in refs.~\cite{Altmannshofer:2017fio,Altmannshofer:2017yso,Aebischer:2019mlg,Altmannshofer:2021qrr}, we consider the following experimental data in the numerical analysis: 1) the branching ratios of the decays $B \to K\mu^+\mu^-$~\cite{LHCb:2014cxe}, $B \to K^*\mu^+\mu^-$~\cite{CDF:2012qwd,LHCb:2014cxe,CMS:2015bcy,LHCb:2016ykl}, $B_s \to \phi \mu^+\mu^-$~\cite{LHCb:2021zwz}, $\Lambda_b \to \Lambda \mu^+ \mu^-$~\cite{LHCb:2015tgy}, $B \to X_s \mu^+\mu^-$~\cite{BaBar:2013qry}, and $B_s \to \mu^+\mu^-$~\cite{LHCb:2021vsc}. 2) the angular distributions in $B \to K \mu^+\mu^-$~\cite{CDF:2012qwd,LHCb:2014auh}, $B \to K^*\mu^+\mu^-$~\cite{CMS:2015bcy,Belle:2016fev,CMS:2017ivg,ATLAS:2018gqc,LHCb:2020lmf,LHCb:2020gog}, $B_s \to \phi \mu^+\mu^-$~\cite{LHCb:2021xxq}, and $\Lambda_b \to \Lambda \mu^+ \mu^-$~\cite{LHCb:2018jna} decays. 3) the LFU ratios $R_{K^{(*)}}$~\cite{BaBar:2012mrf,LHCb:2017avl,Abdesselam:2019wac,BELLE:2019xld,Aaij:2019wad,Aaij:2021vac}. In scenarios I and II, the best fit regions of the $\Zp$ parameters in terms of their real and imaginary parts are shown in figure~\ref{fig:bound:b2smumu}. The best fit values in scenario I (II) read ${\rm Re}\XctL\lmmL/m_\Zp^2=0.0991$ and ${\rm Im}\XctL\lmmL/m_\Zp^2=-0.184$ (${\rm Re}\XctLII\lmmLII/m_\Zp^2=-0.0007$ and ${\rm Im}\XctLII\lmmLII/m_\Zp^2=0.0018$) in unit of $\TeV^{-2}$ with $\text{Pull}_{\rm SM}=4.91\sigma\,(4.96\sigma)$, which is defined as $\sqrt{\Delta\chi^2}$ between the best fit point and the SM value~\cite{Aebischer:2019mlg}. We observe that both the $\Zp$ scenarios I and II allow a good description of the current $b \to s \mu^+ \mu^-$ measurements. This is not surprising, because the $\Zp$ scenarios considered induce NP effects along the direction $\mC_9^\NP=-\mC_{10}^\NP$, which is strongly preferred in the model-independent global fit~\cite{Altmannshofer:2017fio,Altmannshofer:2017yso,Capdevila:2017bsm,Alok:2019ufo,Aebischer:2019mlg,Ciuchini:2020gvn,Alguero:2021anc,Altmannshofer:2021qrr,Geng:2021nhg,Hurth:2021nsi}. In the survived parameter space of the scenarios I and II, the magnitude of $\XctL\lmmL/m_\Zp^2$ and $\XctLII\lmmLII/m_\Zp^2$ are of $\mO(10^{-1})\TeV^{-2}$ and $\mO(10^{-3})\TeV^{-2}$, respectively. In both of these two scenarios, the real parts of $\XctL\lmmL/m_\Zp^2$ and $\XctLII\lmmLII/m_\Zp^2$ are bounded from below, while the imaginary parts are loosely constrained. It is noted that the $b \to s \mu^+ \mu^-$ decays provide the most stringent constraints on the $\Zp$ parameters among all the flavour processes, as will be discussed in the following.

\begin{figure}[t]
  \centering
  \subfigure{\includegraphics[width=0.45\textwidth]{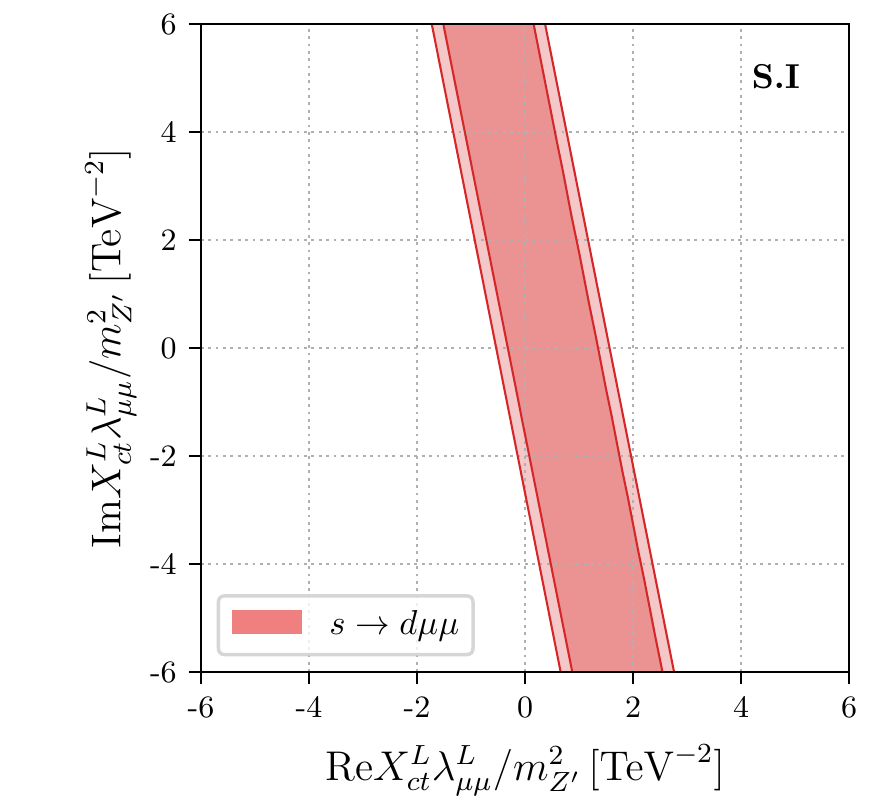}}
  \quad
  \subfigure{\includegraphics[width=0.45\textwidth]{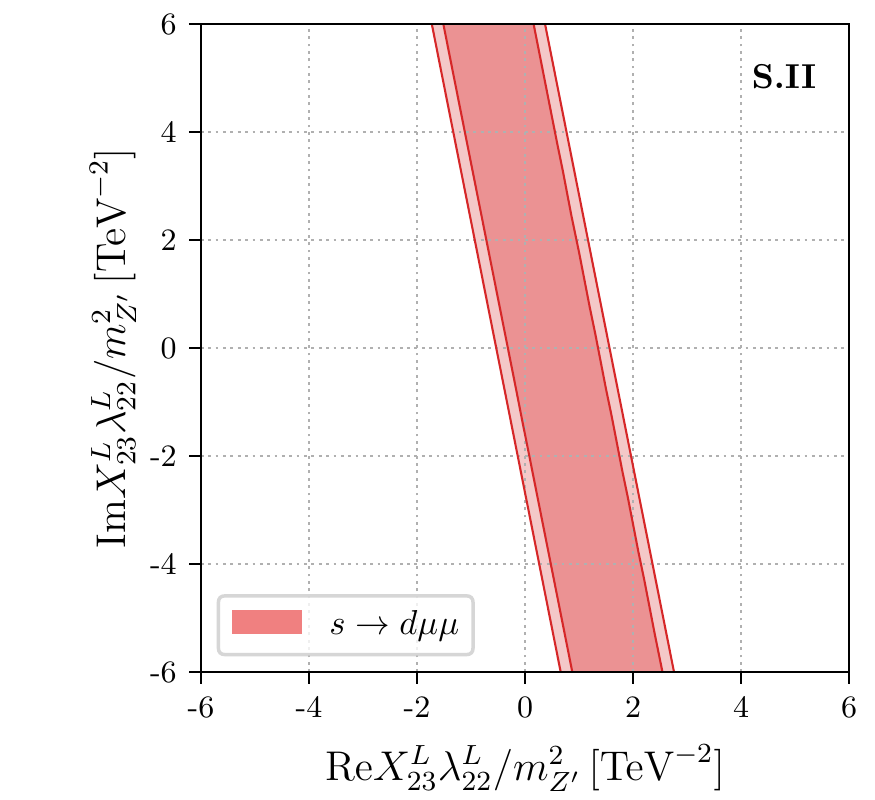}}
  \caption{Same as in figure~\ref{fig:bound:b2smumu} but now under the constraints from the $s \to d \mu^+ \mu^-$ processes.} 
  \label{fig:bound:s2dmumu}
\end{figure}

The rare decays $K_L \to \mu^+ \mu^-$ and $K_S \to \mu^+ \mu^-$ are induced by the $s \to d \mu^+ \mu^-$ transition. In the $\Zp$ scenarios, the constraints from the branching ratios $\mB(K_{L,S} \to \mu^+ \mu^-)_{\rm SD}$ are shown in figure~\ref{fig:bound:s2dmumu}. As discussed in subsection~\ref{sec:b2snunu}, the $\Zp$ boson in scenarios I and II affects the $s \to d \mu^+ \mu^-$ transition via the same penguin diagram. Therefore, the constraints on the $\Zp$ parameters are identical in the two scenarios, as shown in figure~\ref{fig:bound:s2dmumu}. It can also be seen that the bounds on the imaginary parts of $\XctL \lmmL /m_\Zp^2$ and $X_{23}^L \lambda_{22}^L /m_\Zp^2$ are quite weak. In addition, the constraints from the $s \to d \mu^+ \mu^-$ decays, while being relatively weaker compared to that from the $b \to s \mu^+ \mu^-$ decays, are still compatible with the parameter regions required to explain the $b \to s \mu^+ \mu^-$ anomalies shown in figure~\ref{fig:bound:b2smumu}.

\begin{figure}[t]
  \centering
  \subfigure{\includegraphics[width=0.45\textwidth]{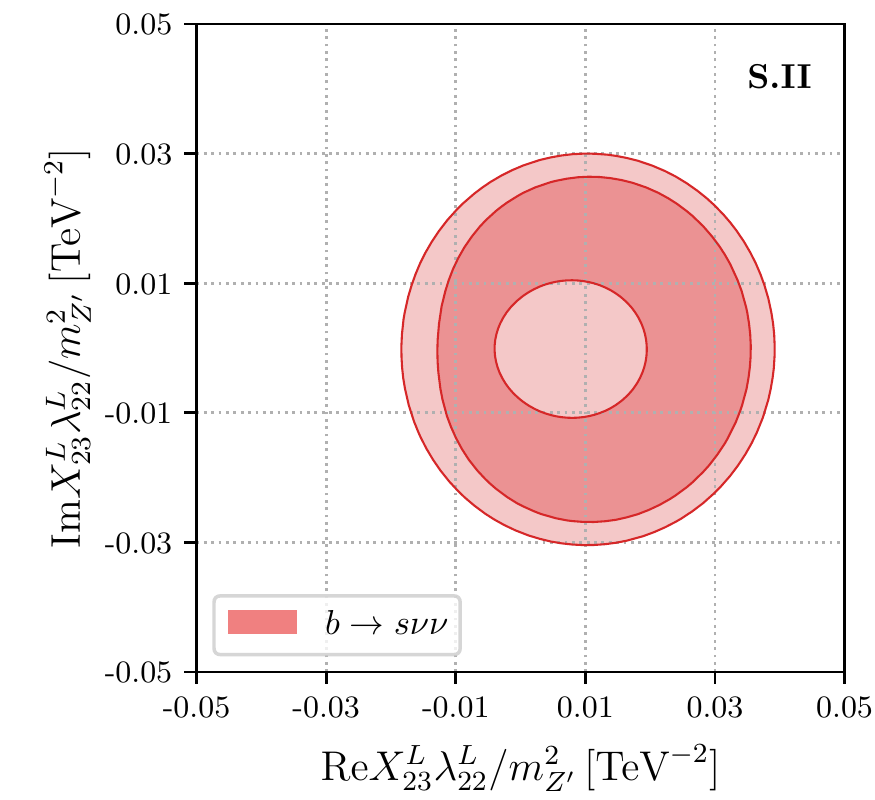}}
  \quad
  \subfigure{\includegraphics[width=0.45\textwidth]{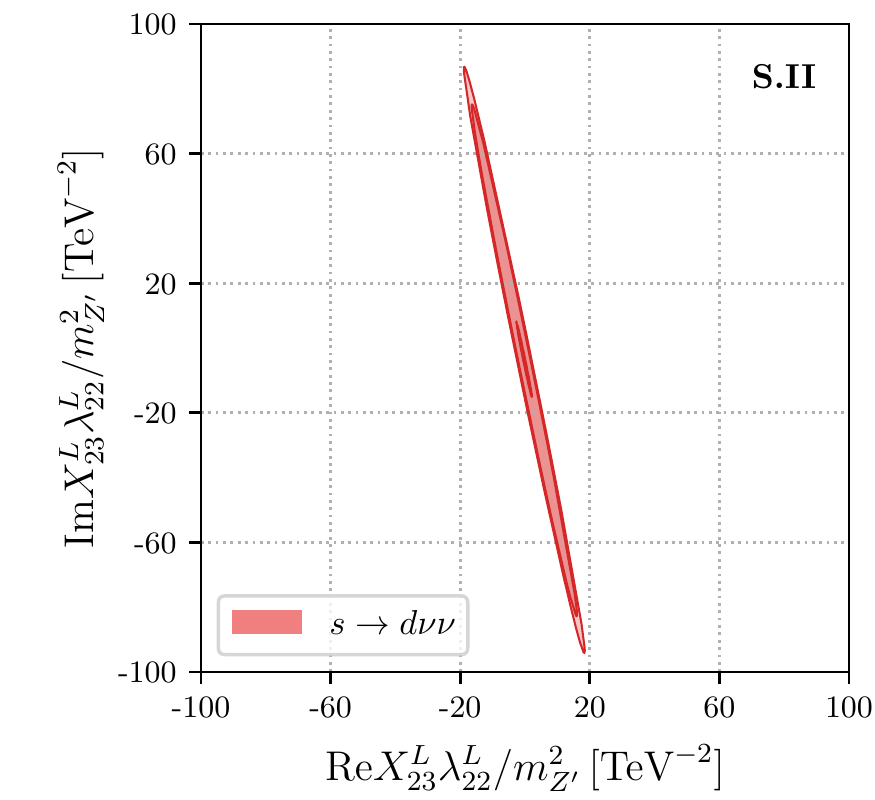}}
  \caption{Allowed regions of the real and imaginary parts of $X_{23}^L\lambda^{L}_{22}/m_\Zp^2$ by the $b \to s \nu \bar\nu$ (left) and $s \to d \nu\bar\nu$ processes (right) in scenario II. The colour captions are the same as in figure~\ref{fig:bound:b2smumu}.}
  \label{fig:bound:b2snunu}
\end{figure}

Since the $\Zp$ boson does not couple directly to the neutrinos in the $\Zp$ scenario I, the $b \to s \nu \bar\nu$ and $s \to d \nu \bar \nu$ decays are not relevant for this scenario. In the $\Zp$ scenario II, on the other hand, the $\Zp$ boson affects the $b \to s \nu\bar\nu$ and $s \to d \nu \bar\nu$ transitions at the tree and the one-loop level, respectively. In order to constrain the $\Zp$ parameters, we consider five $b \to s \nu \bar \nu$ processes (i.e., the decays $B \to X_s \nu \bar\nu$, $B^0 \to K^{(*)0} \nu \bar\nu$, and $B^+ \to K^{(*)+} \nu \bar\nu$), and two $s \to d \nu \bar\nu$ processes (i.e., the decays $K^+ \to \pi^+ \nu \bar\nu$ and $K_L \to \pi^0 \nu \bar\nu$). We show in figure~\ref{fig:bound:b2snunu} the allowed regions of the real and imaginary parts of $X_{23}^L \lambda_{22}^L/m_\Zp^2$ by the branching ratios of these decays. It can be seen that the bound from the $s \to d \nu \bar\nu$ decays is much weaker than that from the $b \to s \nu \bar\nu$ transitions. This can be understood by the fact that the $\Zp$ contributions to the former is suppressed by $\mO(\lambda^3)$ while its contributions to the latter do not suffer any CKM suppression. In addition, the constraints from the $b \to s \nu \bar\nu$ and $s \to d \nu \bar \nu$ decays are consistent with the best fit region from the $b \to s \mu^+ \mu^-$ processes.

\begin{figure}[t]
  \centering
  \includegraphics[width=0.45\textwidth]{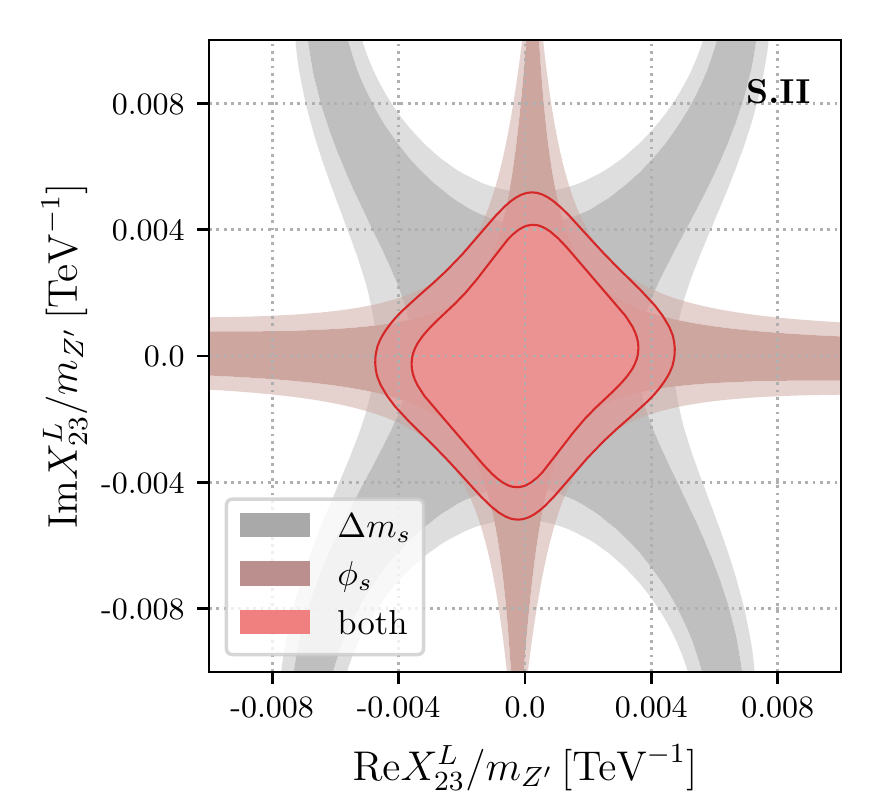}
  \caption{Allowed regions in the plane $\big (\re\XctLII/m_\Zp, \im\XctLII/m_\Zp \big)$ by the $B_s - \bar B_s$ mixing in the $\Zp$ scenario II. The dark and the light regions correspond to the $68\%$ and the $95\%$ CL, respectively.}
  \label{fig:bound:Bs-mixing}
\end{figure}

In the $\Zp$ scenario I, as explained in subsection~\ref{sec:mixing}, the $B_s - \bar B_s$ mixing cannot provide any relevant constraint on the $\Zp$ parameters. In the $\Zp$ scenario II, the $\Zp$ contribution is proportional to the product $(\XctLII)^2/m_\Zp^2$ due to the tree-level $\Zp$ exchange. After taking into account the constraints from the mass difference $\Delta m_s$ and the CP-violating observable $S_{\psi\phi}$, we obtain the allowed regions in the plane $\big( \re\XctLII/m_\Zp, \im\XctLII/m_\Zp \big)$, which are shown in figure~\ref{fig:bound:Bs-mixing}. It can be seen that the mass difference $\Delta m_s$ provides a strong bound on $\re\XctL/m_\Zp$. For $\im\XctLII/m_\Zp$, the bound arises mainly from the CP-violating observable $S_{\psi\phi}$ and is weaker compared to the one on $\re\XctLII/m_\Zp$, due to the currently larger experimental uncertainty of $S_{\psi\phi}$. The phase of the maximum value of $|\XctLII|/m_\Zp$ also matches the CKM phase $\beta_s$~\cite{Charles:2004jd}.

Summarizing the numerical analysis made above, we can see that the strongest constraints are provided by the $b \to s \mu^+ \mu^-$ processes in the $\Zp$ scenario I. Therefore, after considering all the flavour processes, we can obtain the combined allowed regions from figure~\ref{fig:bound:b2smumu}, with the numerical results given by
\begin{align}
  0.04 < \frac{\re\XctL\lmmL}{m_\Zp^2} < 0.21 \TeV^{-2},
  \quad\quad
  -0.44 < \frac{\im\XctL\lmmL}{m_\Zp^2} < 0.41 \TeV^{-2},
\end{align}
at the $95\%$ CL. The above lower bound on $\re\XctL\lmmL/m_\Zp^2$, together with the parameter space chosen in eq.~(\ref{eq:parameter space}), implies an upper bound on the $\Zp$ mass, $m_\Zp<9.0\TeV$.

Turning to the $\Zp$ scenario II, the $b \to s \mu^+ \mu^-$ processes also provide the most stringent constraints on the $\Zp$ parameter, as in scenario I. Therefore, after considering all the $B$- and $K$-meson FCNC decays, the combined constraints can be obtained from figure~\ref{fig:bound:b2smumu}, and numerically
\begin{align}
  -0.0015 < \frac{\re\XctLII\lmmLII}{m_\Zp^2} < -0.0003 \TeV^{-2},
	\quad\quad
  -0.0029 < \frac{\im\XctLII\lmmLII}{m_\Zp^2} < +0.0031 \TeV^{-2},
\end{align}
at the $95\%$ CL. However, different from the scenario I, the $ B_s - \bar B_s$ mixing also provides an independent bound on the parameter product $\XctLII/m_\Zp$. Numerically, the combined constraints shown in figure~\ref{fig:bound:Bs-mixing} correspond to
\begin{align}
  -0.004 < \frac{\re\XctLII}{m_\Zp} < +0.004 \TeV^{-1},
  \quad 
  -0.005 < \frac{\im\XctLII}{m_\Zp} < +0.005 \TeV^{-1},
\end{align}
at the $95\%$ CL. In the case of $\lmmLII/m_\Zp \sim \mO(1)\TeV^{-1}$, constraints on $\XctLII/m_\Zp$ from the $b \to s \mu^+ \mu^-$ processes and the $B_s - \bar B_s$ mixing are of the same order. Furthermore, since $\re\XctLII\lmmLII/m_\Zp^2$ is lower bounded by the $b \to s \mu^+ \mu^-$ processes, a lower bound, $\lmmLII/m_\Zp > 0.07\TeV^{-1}$, can be derived in order to simultaneously satisfy the constraints from the $B_s - \bar B_s$ mixing. Therefore, the $\Zp$ couplings in scenario II exhibit the hierarchy $\lmmLII \gg \XctLII$. Finally, our numerical analysis has shown that the main constraints in scenario II are obtained from the processes affected by the tree-level $\Zp$ contributions. Therefore, the flavour phenomenology of the $\Zp$ scenario II is almost identical to that of the minimal $\Zp$ scenario discussed in refs.~\cite{Gauld:2013qba,DiChiara:2017cjq}.

\subsection{Collider constraints}
\label{sec:num:high energy}

In this subsection, we discuss the collider constraints on the $\Zp$ scenarios in the case of both $m_\Zp > m_t$ and $m_\Zp < m_t$. For recent collider studies of similar $\Zp$ scenarios with top-FCNC couplings, we refer to  refs.~\cite{Hou:2017ozb,Cho:2019stk,Alvarez:2020ffi}.

\subsubsection{$m_\Zp > m_t$}
\label{sec:num:high energy:heavy Zp}

In the case of $m_\Zp > m_t$, constraints on the $\Zp$ parameters could be obtained from the decay $t \to c \mu^+ \mu^-$. However, the current LHC searches for the decay $t \to c \mu^+ \mu^-$ have been only performed at the $Z$ peak, with $|m_{\mu^+ \mu^-} - m_Z| < 15\,\GeV$, and interpreted as a bound on $\mB(t \to c Z)$~\cite{CMS:2017wcz,CMS-PAS-TOP-17-017,ATLAS:2018zsq,ATLAS:2021stq}. No dedicated searches for non-resonant (outside the $Z$ peak) $t \to c \mu^+ \mu^-$ decays have been performed yet. In ref.~\cite{Durieux:2014xla}, an upper bound on $\mB(t \to c \mu^+ \mu^-)$ was estimated by using the experimental bounds on $\mB(t \to c Z)$ and $\mB(Z \to \mu^+ \mu^-)=3.37\%$~\cite{PDG:2020}. With such an approach, an upper bound $\mB(t \to c \mu^+ \mu^-) < 4.4 \times 10^{-6}$ can be derived from the current bound $\mB(t \to c Z)<1.3 \times 10^{-4}$ set by the ATLAS experiment with an integrated luminosity of $139\fb^{-1}$ at $13\TeV$~\cite{ATLAS:2021stq}. In this way, constraints on the $\Zp$ parameters can be derived from the bound $\mB(t \to c \mu^+ \mu^-) < 4.4 \times 10^{-6}$. However, we find that such a constraint is at least one order of magnitude weaker than that obtained from the low-energy flavour processes discussed in the last subsection. 

Within the effective field theory (EFT) framework, by using different signal regions of the LHC searches for the rare FCNC decay $t \to c Z$, an improved approach has been developed in ref.~\cite{Chala:2018agk}. However, in the case when $m_\Zp$ is not far from $m_t$, the framework with four-fermion effective operators is not appropriate to describe the $\Zp$ contributions to the decay $t \to c \mu^+ \mu^-$.

\subsubsection{$m_\Zp < m_t$}
\label{sec:num:high energy:heavy Zp}

\begin{figure}[t]
	\centering
	\includegraphics[width=0.45\textwidth]{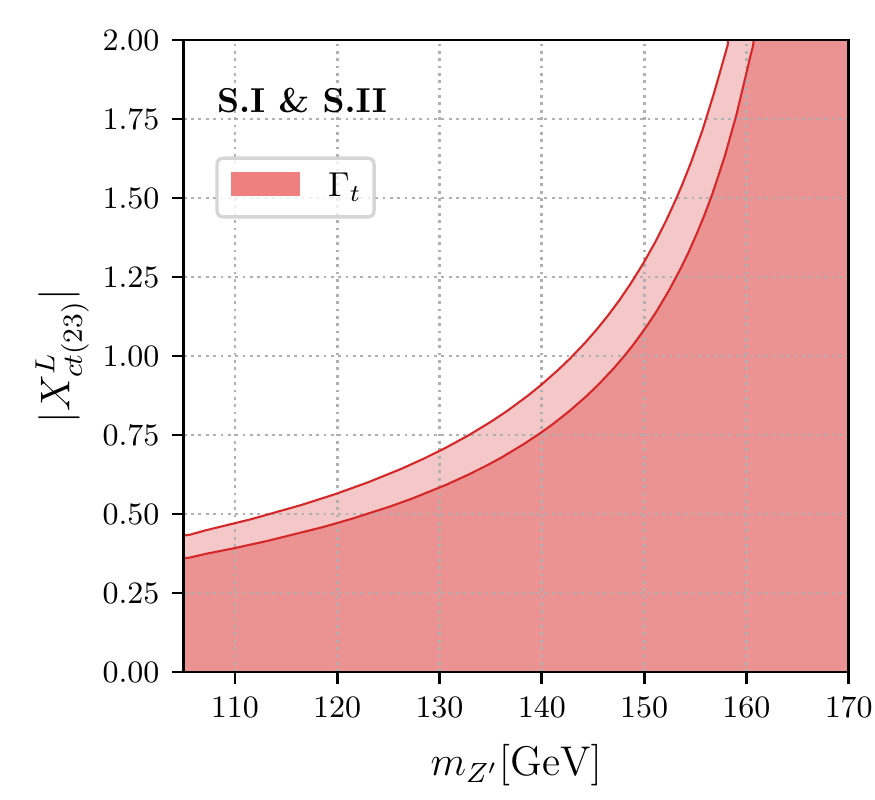}
	\caption{Allowed regions in the plane $(\lvert X_{ct(23)}^L\rvert, m_\Zp)$ from the top-quark width $\Gamma_t$ in the $\Zp$ scenario I (II). The colour captions are the same as in figure~\ref{fig:bound:b2smumu}.} 
	\label{fig:bound:top width}
\end{figure}

In the case of $m_\Zp < m_t$, the $t \to c \mu^+ \mu^-$ decay involves the resonant $\Zp$ contribution, and is expected to provide strong constraints on the $\Zp$ parameters. However, similar to the case of $m_\Zp > m_t$, no experimental searches for the $t \to c \mu^+ \mu^-$ decay mediated by the $\Zp$ resonance has been performed yet. Detailed analysis of the signal shape could be used to derive constraints from the current $\mB(t \to c Z)$ bound, as performed in ref.~\cite{Chala:2018agk}. Especially, for $\lvert m_\Zp - m_Z \rvert < 15\,\GeV$, the signal regions of the $t \to c Z$ and $t \to c \Zp$ decays can be largely overlapped, and stringent constraints on the $\Zp$ parameters could be therefore derived. We leave these possibilities for further work, and concentrate on the mass region $105\GeV < m_\Zp < m_t$ to avoid the potentially strong bound from $\mB(t \to c Z)$~\cite{ATLAS:2021stq}.

As discussed in subsection~\ref{sec:t2cmumu:light Zp}, in the case of $m_\Zp < m_t$, the decay $t \to c \Zp$ can contribute to the top-quark width. Compared to the SM prediction $\Gamma_t^\SM=1.3\,\GeV$~\cite{Gao:2012ja}, the current measurement $\Gamma_t=(1.42_{-0.15}^{+0.19})\,\GeV$~\cite{PDG:2020} leaves $\mO(20\%)$ room for the $\Zp$ effects. We show in figure~\ref{fig:bound:top width} the allowed regions in the plane $(\lvert X_{ct(23)}^L\rvert, m_\Zp)$ in the scenario I (II). For scenario I, it is noted that the top-quark width provides a unique bound on the coupling $\XctL$. For scenario II, the bound on $\XctLII$ is much weaker than that obtained from the $B_s - \bar B_s$ mixing.

\subsection{Predictions}

\begin{figure}[t]
	\centering
	\includegraphics[width=0.45\textwidth]{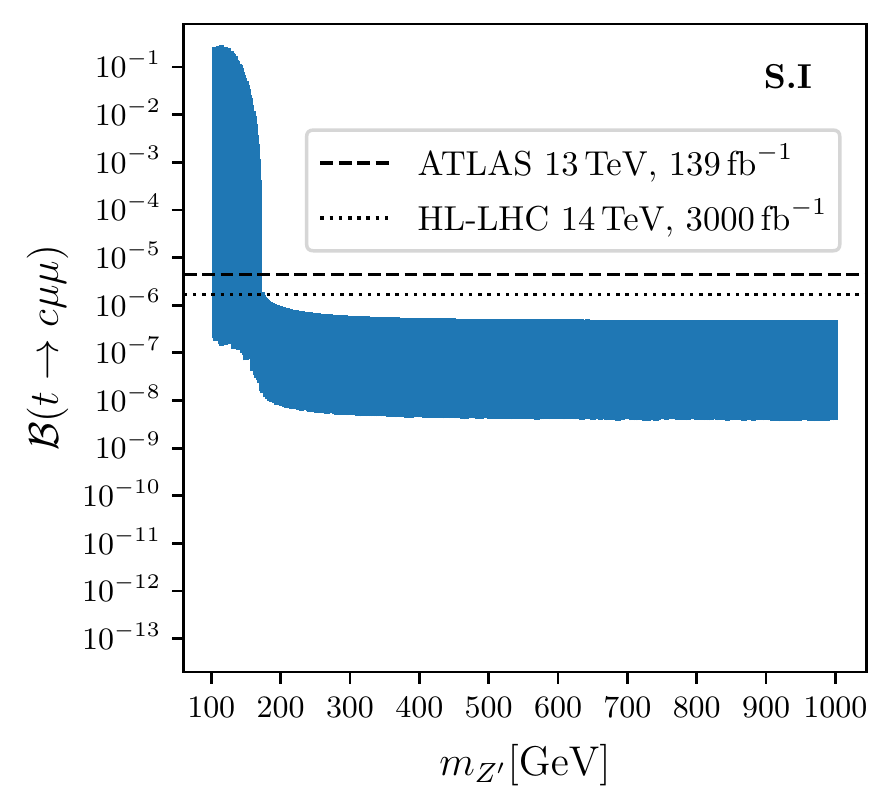}
	\qquad
	\includegraphics[width=0.45\textwidth]{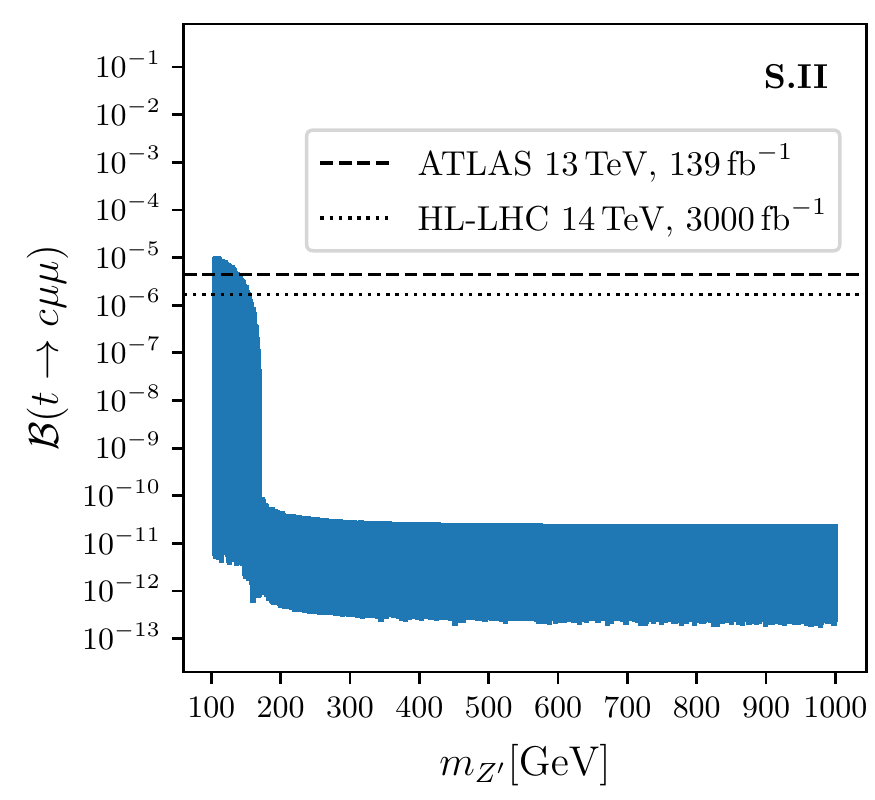} 	
	\caption{The branching ratio of $t \to c \mu^+ \mu^-$ decay as a function of $m_\Zp$ in the $\Zp$ scenarios I (left) and II (right). The dark regions are allowed at the $95\%$ CL.}
	\label{fig:prediction:t2cmumu}
\end{figure}

As mentioned in the last subsection, experimental searches for the $t \to c \mu^+ \mu^-$ decay off the $Z$ pole have not been performed yet. Using the allowed parameter space derived in the global fit, we can make prediction on the branching ratio $\mB(t \to c \mu^+ \mu^-)$ in the two $\Zp$ scenarios as a function of the $\Zp$ mass, which is shown in figure~\ref{fig:prediction:t2cmumu}. Although there are several studies of the expected sensitivities to the $t \to c Z$ decay at the LHC~\cite{ATLAS:2019pcn} and other future colliders~\cite{Liu:2020bem,Behera:2018ryv,Khatibi:2021phr}, to our knowledge, detailed analysis of the $t \to c \mu^+ \mu^-$ decay at the LHC has not been performed yet. In order to estimate the current (future) experimental sensitivity to $\mB(t \to c \mu^+ \mu^-)$, we adopt as a benchmark the product of $\mB(Z \to \mu^+ \mu^-)=3.37\%$~\cite{PDG:2020} and the current (future projected) experimental limit on $\mB(t \to c Z)$. With the current ATLAS limit $\mB(t \to c Z)<1.3 \times 10^{-4}$ based on the full Run-2 data~\cite{ATLAS:2021stq} and the future expected sensitivity $5 \times 10^{-5}$ at the HL-LHC with $3\,\mathrm{ab}^{-1}$~\cite{ATLAS:2019pcn}, the current and the future sensitivity to $\mB(t \to c \mu^+ \mu^-)$ are estimated to be $4.4 \times 10^{-6}$ and $1.7 \times 10^{-6}$ respectively, as shown by the dashed and the dotted line in figure~\ref{fig:prediction:t2cmumu}. Here we make the following two observations:
\begin{itemize}
\item In the case of $m_\Zp < m_t$, the branching ratio of $t \to c \mu^+ \mu^-$ decay in the two $\Zp$ scenarios is strongly enhanced by the resonance effect, which is several orders of magnitude higher than the one in the case of $m_\Zp > m_t $. In the mass region $105\GeV < m_\Zp < m_t$, the predicted $\mB ( t \to c \mu^+ \mu^-)$ in scenario I (II) is higher (lower) than the current and future estimated bounds. Therefore, experimental searches for the $t \to c \mu^+ \mu^-$ decay are expected to probe a $\Zp$ mass window of $[105\GeV, m_t]$.

\item In the case of $m_\Zp > m_t$, the predicted $\mB(t \to c \mu^+ \mu^-)$ in scenario I is  compatible with the estimated sensitivities, while the one in scenario II is several orders of magnitude lower the estimated future sensitivity. Therefore, direct searches for the decay in scenario II with a heavy $\Zp$ boson could be very challenging for the LHC and its high-luminosity upgrade.
\end{itemize}

\begin{figure}[t]
	\centering
	\includegraphics[width=0.45\textwidth]{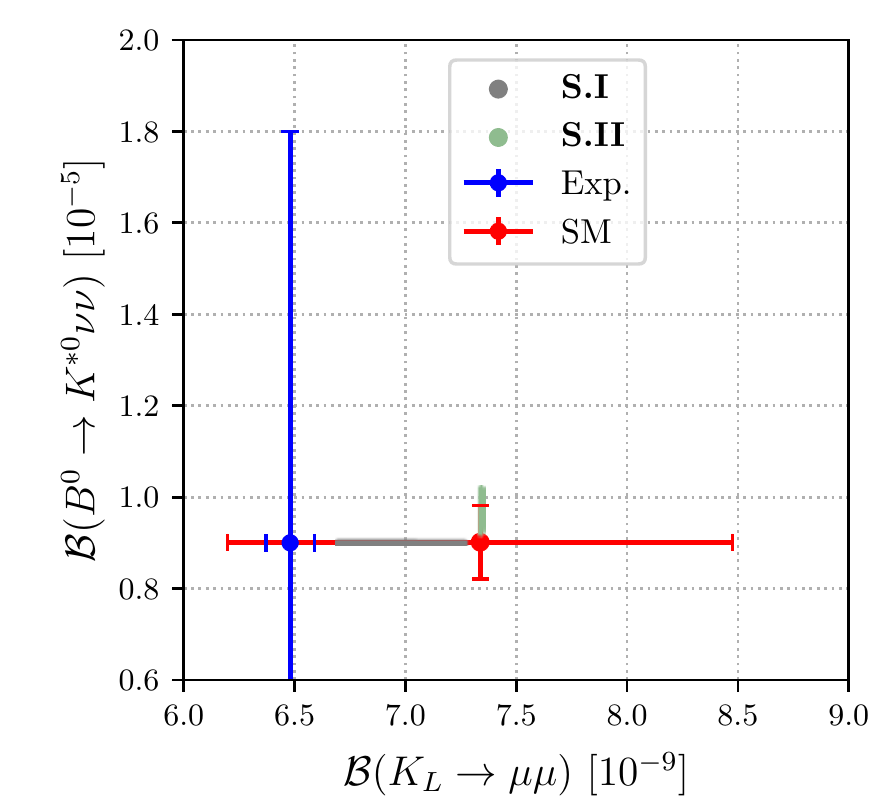}
	\qquad
	\includegraphics[width=0.45\textwidth]{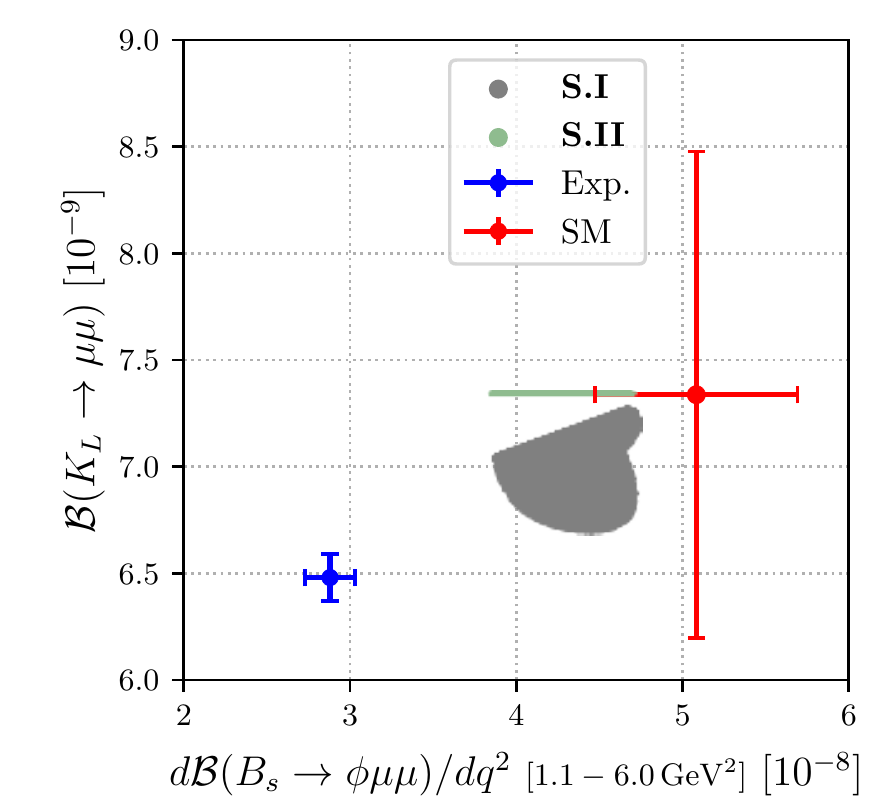}
	\\
	\includegraphics[width=0.45\textwidth]{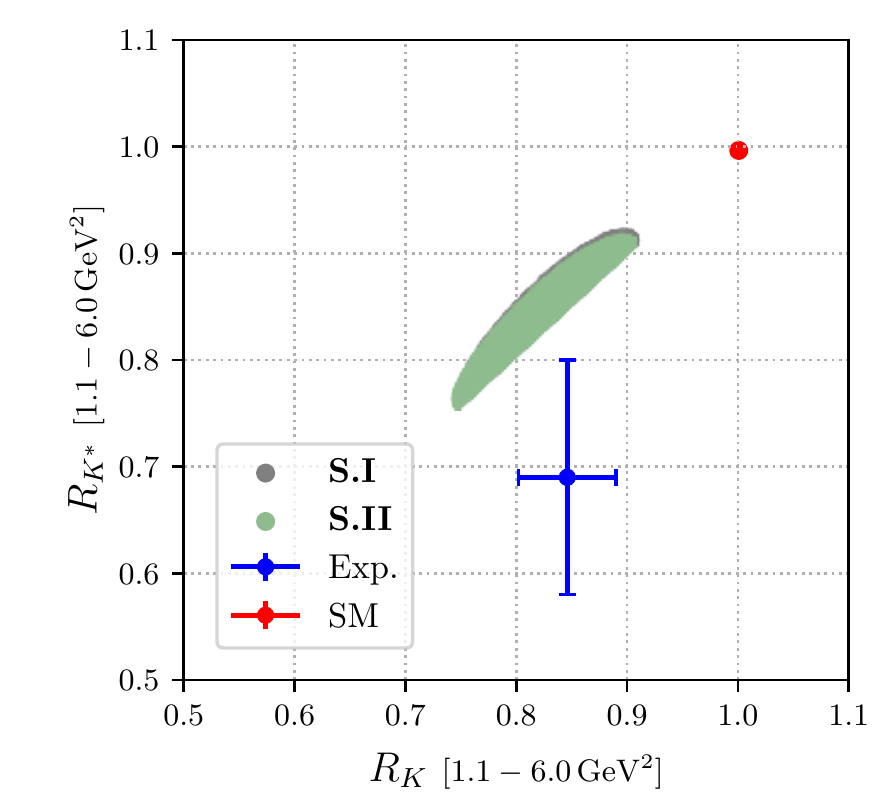}
	\qquad
	\includegraphics[width=0.45\textwidth]{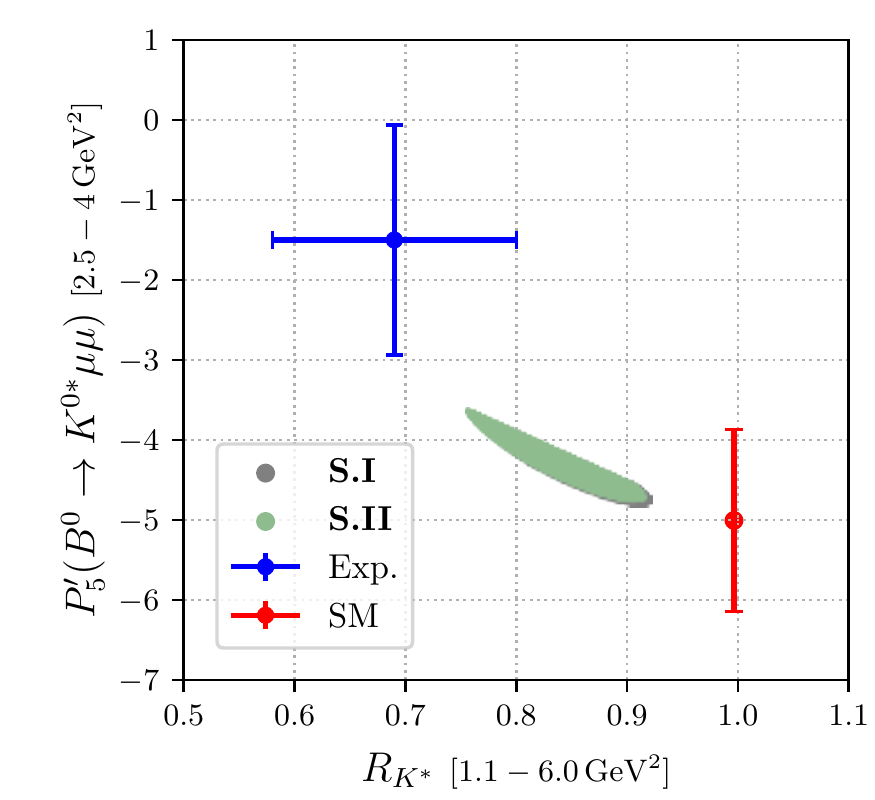}
	\caption{Correlations between the observables of $B$- and $K$-meson decays in the two $\Zp$ scenarios. The $1\sigma$ range of the experimental measurements and the SM predictions are shown, except the data on $\mB(B^0 \to K^{*0} \nu\nu)$, which corresponds to the $90\%$ CL limit.}
	\label{fig:correlation}
\end{figure}

Besides the $t \to c \mu^+ \mu^-$ decay, which should be searched in $ pp \to t \bar t$ production, the associated production $pp \to t\Zp$ could also be considered and may provide more relevant constraints for a heavy $\Zp$ boson. Recently, the associated production of a single top with dileptons has been investigated in the EFT framework~\cite{Afik:2021jjh}, and a $95\%$ CL bound on the scale of the effective operator $(\bar t \gamma^\mu P_L c)(\bar\mu\gamma_\mu P_L \mu)$, $\Lambda_{tc\mu\mu}=1.1\,(1.5)\,\TeV$, has been obtained at the LHC with an integrated luminosity of $140\,(3000)\fb^{-1}$. In the $\Zp$ scenario I (II) with $m_\Zp=1\,\TeV$, the $95\%$ CL allowed parameter space in our global fit results in $1.4 \,(17.0) < \Lambda_{tc\mu\mu} < 4.8\,(55.2)\TeV$. This implies that the $\Zp$ scenario I reaches the sensitivity of the HL-LHC even for a TeV-scale $\Zp$ boson.\footnote{In the $\Zp$ scenario II, the $b\bar{s}\Zp$ coupling makes the $pp \to \mu^+ \mu^-$ process sensitive to the $\Zp$ boson. For more details, we refer to ref.~\cite{Allanach:2018odd}.} Detailed collider simulation is still necessary to obtain more reliable estimation of the experimental sensitivity, which will be our future work. We also encourage our experimental colleagues to carry out relevant searches for such a $\Zp$ boson.

An important feature of our $\Zp$ scenarios is that the $\Zp$ contributions to all the FCNC decays are controlled by the same product, $\XctL\lmmL/m_\Zp^2$ in scenario I or $\XctLII\lmmLII/m_\Zp^2$ in scenario II. Therefore, all the $B$- and $K$-meson processes are strongly correlated. As an illustration, we show in figure~\ref{fig:correlation} the correlations among some observables. From these plots, we make the following two observations:
\begin{itemize}
\item The branching ratio $\mB(K_L \to \mu^+ \mu^-)$ is suppressed by the $\Zp$ effects in scenario I. However, its SM value remain almost unchanged in scenario II, due to the strongly constrained $\Zp$ couplings. For the $B^0 \to K^{*0}\mu^+\mu^-$ decay, the situation is opposite: the branching ratio is enhanced in scenario II, while remaining unchanged in scenario I. These correlations can be used to distinguish the two $\Zp$ scenarios. In addition, the $\Zp$ effects on the branching ratios of $K_S \to \mu^+ \mu^-$, $K_L \to \pi \nu \bar\nu$ and $K^+ \to \pi^+ \nu \bar\nu$ decays are all found to be negligible in the two scenarios and will be not shown here.
  
\item The correlations between $R_K^{[1.1,\,6.0]}$ and $R_{K^*}^{[1.1,\,6.0]}$, which are the LFU ratios in the $q^2$ bin $[1.1,\,6.0]\,\GeV^2$, as well as the correlations between $R_{K^*}^{[1.1,\,6.0]}$ and $P_5^\prime$, in the two $\Zp$ scenarios are almost identical. The reason is that the best fit regions for the Wilson coefficients $\mC_9 $ and $\mC_{10}$ are numerically almost the same in the two scenarios. As expected, the $\Zp$ effects make $R_{K^{(*)}}^{[1.1,\,6.0]}$, $P_5^\prime$, and $\mB(B_s \to \phi \mu^+ \mu^-)$ move closer to the experimental measurements.
\end{itemize}
With high-precision measurements at the HL-LHC and Belle II~\cite{Belle-II:2018jsg}, these interesting correlation could provide further insights into our $\Zp$ scenarios.

\begin{figure}[t]
  \centering
  \includegraphics[width=0.45\textwidth]{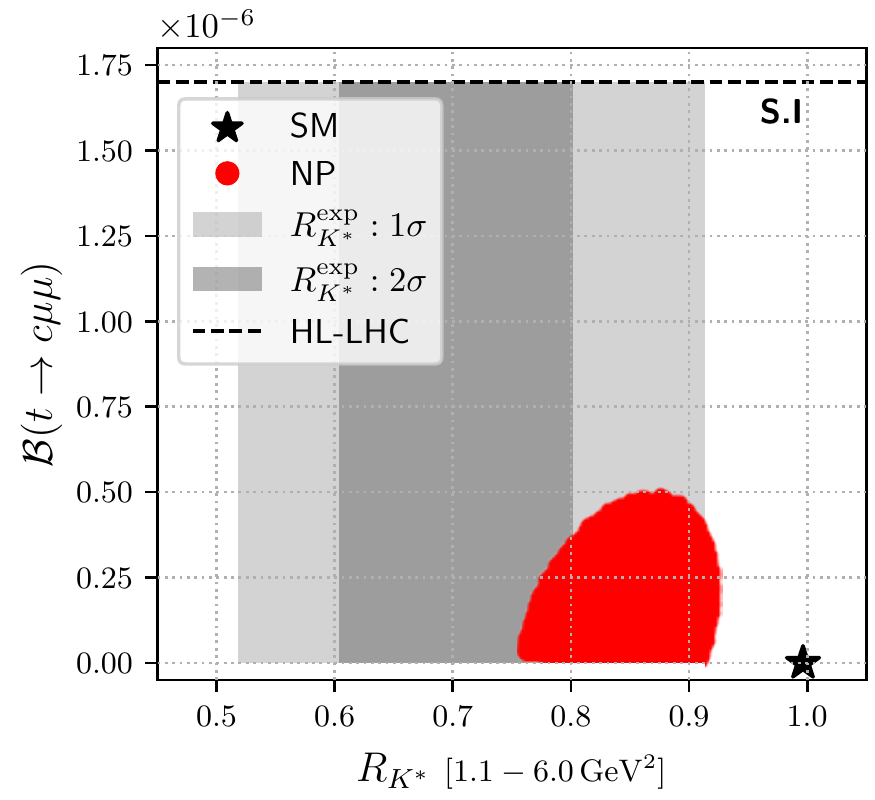}
  \qquad
  \includegraphics[width=0.45\textwidth]{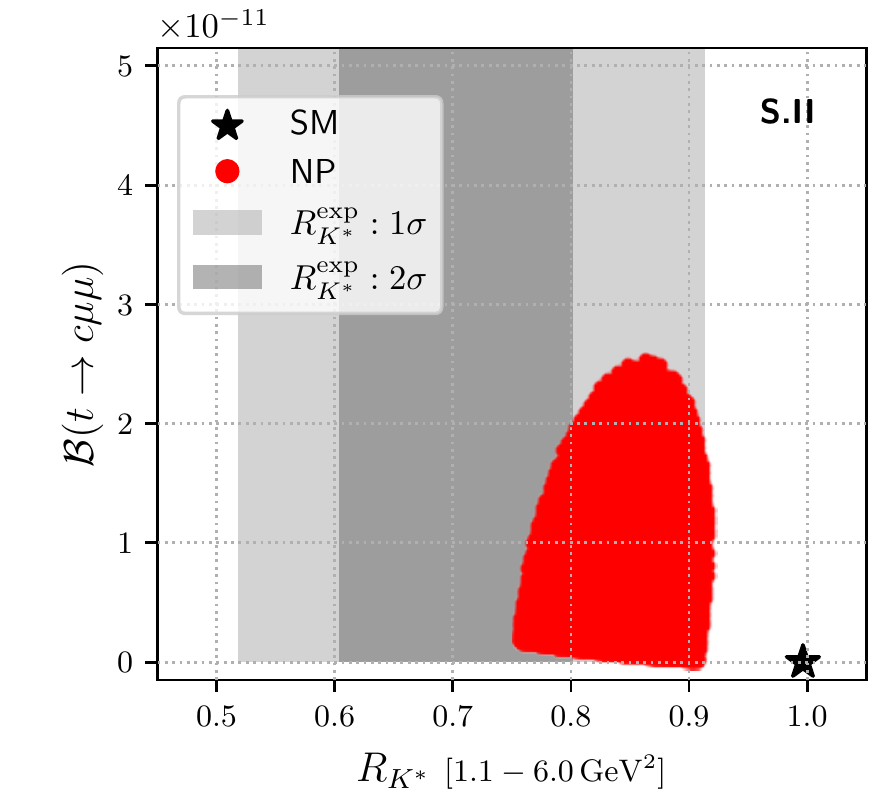}
  \caption{Correlation between $R_{K^*}$ and $\mB(t\to c\mu^+\mu^-)$ in the $ Z^{\prime} $ scenarios I (left) and II (right) for $m_\Zp=1\,\TeV$ with real (gray) and complex (red) couplings.} 
  \label{fig:correlation:RKstar:t2cmumu}
\end{figure}

In the case of $m_\Zp \gg m_t$, the $\Zp$ contributions to both the $t \to c \mu^+ \mu^-$ and $b \to s \mu^+ \mu^-$ processes are controlled by the same products, $\XctL\lmmL/m_\Zp^2$ in scenario I or $\XctLII\lmmLII/m_\Zp^2$ in scenario II, since they are described by the effective $t\bar{c} \mu^+ \mu^-$ operator. Thus, we show in figure~\ref{fig:correlation:RKstar:t2cmumu} the correlation between $R_{K^*}^{[1.1,\,6.0]}$ and $\mB(t \to c \mu^+ \mu^-)$ for $m_\Zp = 1\,\TeV$. It can be seen that $\mB(t \to c \mu^+ \mu^-)$ approaches to its maximum value when $R_{K^*}^{[1.1,\,6.0]} \approx 0.85$. The predicted upper limit is of the same order as the estimated future LHC sensitivity in scenario I, while being quite below the estimated sensitivity in scenario II.

\section{Conclusions}
\label{sec:conclusion}

Motivated by the recent anomalies in the $b \to s \ell^+ \ell^-$ transitions, we have considered a phenomenological $\Zp$ scenario (denoted as scenario I), in which a heavy $\Zp$ boson couples only to $t \bar c$ and $\mu^+ \mu^-$ with left-handed chirality. The $\Zp$ effects on the $b \to s \mu^+ \mu^-$ processes automatically induce opposite contributions to the effective operators $\mO_9$ and $\mO_{10}$, which is favored by the model-independent analyses of the $b \to s \ell^+ \ell^-$ anomalies.

We have performed a global fit to all relevant experimental data. It is found that such a minimal $\Zp$ scenario can address the current $b \to s \ell^+ \ell^-$ anomalies, while simultaneously satisfying other flavour and collider constraints. The mass of the $\Zp$ boson can be less than $1\,\TeV$. In the region $105\GeV<m_\Zp < m_t$, the $t \to c \mu^+ \mu^-$ decay is significantly enhanced by resonance effects and can serve as a sensitive probe of such a $\Zp$ boson. As an important feature of this scenario, all the low-energy flavour observables are controlled by the product $\XctL\lmmL/m_\Zp^2$. We have found interesting correlations among the various flavour observables, which could provide further insights into this scenario.

We have also considered an extended scenario (denoted as scenario II), in which the $\Zp$ boson interacts with the $SU_L(2)$ fermion doublets with analogous couplings as in scenario I. Due to tree-level $\Zp$ contributions, the $\Zp$ couplings suffer from severe constraints from the $b \to s \ell^+ \ell^-$ processes and the $B_s - \bar B_s$ mixing, which makes the collider signals of $t \to c \mu^+ \mu^-$ decay below the estimated sensitivity at the HL-LHC. The correlations between flavour observables in scenario II are different from that observed in scenario I, and can be therefore used to distinguish between the two scenarios.

Our scenarios can be modified by replacing $ t \bar c\Zp$ with $t \bar u\Zp$ coupling. Then the $K$-meson rare decays could become more relevant due to the large CKM factors involved, and the $t\Zp$ associated production at the LHC may play crucial role in searching for such a $\Zp$ boson. A right-handed $\mu^+\mu^-\Zp$ coupling can also be taken into account to simultaneously accommodate the $(g-2)_\mu$ anomaly. In addition, if the $\mu^+ \mu^- \Zp$ coupling is replaced by the $e^+ e^- \Zp$ coupling, the direct production of the $\Zp$ boson at $e^+e^-$ collider should then be taken into account. It is noted that, although our $\Zp$ scenarios are not realistic, our studies have shown that a vector boson with top-quark FCNC interactions can explain the current $b \to s \ell^+ \ell^-$ anomalies and may provide new avenues for model buildings. We also encourage our experimental colleagues to carry out relevant searches for such a $\Zp$ boson at the LHC and its high-luminosity upgrade.

\section*{Acknowledgements}
This work is supported by the National Natural Science Foundation of China under Grant Nos. 12135006, 12075097, 11675061, 11775092, and 11805077, as well as by the self-determined research funds of CCNU from the colleges’ basic research and operation of MOE under Grant Nos. CCNU19TD012 and CCNU20TS007. XY is also supported in part by the Startup Research Funding from CCNU.

\bibliographystyle{JHEP}

\begingroup
\setlength{\bibsep}{5pt}
\setstretch{1.2}
\bibliography{ref}
\endgroup

\end{document}